\documentstyle[preprint,graphicx,eqsecnum,tighten,floats,aps]{revtex}
\makeatletter

\def\M{{\cal M}}
\def\D{{\Delta}}
\def\PD{\widehat\Delta}
\def\SD{S_\Delta}
\def\Si{S_\infty}

\def\PDp#1{\widehat{#1}}

\def\hc{\omega}
\def\te{\,{}^2\!\epsilon}

\def\AD{a_\Delta}
\def\RD{R\Delta}
\def\JD{J_\Delta}
\def\MD{M_\Delta}
\def\RD{R_\Delta}

\def\emA{{\bf A}}
\def\emF{{\bf F}}
\def\emQD{{\bf Q}_\Delta}

\def\sa{{\cal S}}
\def\psa{\hat{\cal S}}
\def\sai{{\cal I}}
\def\pka{{\cal K}}

\def\L{{\cal L}}
\def\G{{{\rm I}\!\Gamma}}

\def\ed{{\rm d}}
\def\dd{{\rm d\!I}}
\def\dual{\star}

\def\otop#1{\vbox{\mathsurround=0pt 
  \ialign{##\crcr$\hss\scriptstyle\circ\hss$\crcr
    \noalign{\kern1pt\nointerlineskip}$\displaystyle{#1}$\crcr}}
  \vphantom{#1}}

\def\eqhat{\mathrel{\widehat\mathalpha{=}}}

\def\gen{\cal P}

\def\eqref#1{(\ref{#1})}

\def\ImPt#1{{\rm Im} \left[ #1 \right]}

\def\Lie{{\cal L}}

\def\Re{{\rm R}}

\def\vins{\mathbin{\dimen0=\ht\strutbox  \divide\dimen0 by 2
  \hbox{\vbox{\hrule width\dimen0}\hskip-0.4pt\vrule height\dimen0}}\,}

\def\Tr#1{{\rm Tr} \left[ #1 \right]}

\def\pback#1{{
\mathchoice{\StemPullBack{#1}{\leftarrowfill}}
	 {\StemPullBack{#1}{\leftarrowfill}}
             {\IndxPullBack{#1}{\leftarrowfill}}
		 {\IndxPullBack{#1}{\leftarrowfill}}}\vphantom{#1}}

\newcommand{\StemPullBack}[2]{
  \vtop{\mathsurround=0pt  
  \ialign{##\crcr$\textstyle{#1}\strut$\crcr
    \noalign{\kern-0.4ex\nointerlineskip}{\tiny#2}\crcr}}}

\newcommand{\IndxPullBack}[2]{
  \vtop{\mathsurround=0pt  
  \ialign{##\crcr\hfil$\scriptstyle{#1}$\hfil\crcr
    \noalign{\kern+0.4ex\nointerlineskip}{\tiny#2}\crcr}}}

\begin{document}

\title{Mechanics of Rotating Isolated Horizons}
\author {Abhay Ashtekar${}^{1,4}$\thanks{E-mail: 
ashtekar@gravity.phys.psu.edu},
Christopher Beetle${}^{1,2}$\thanks{E-mail: beetle@physics.utah.edu}, 
and Jerzy Lewandowski${}^{3,4}$\thanks{E-mail:jerzy.lewandowski@fuw.edu.pl} } 
\address{{1.} Center for Gravitational Physics and Geometry\\
Physics Department, Penn State, University Park, PA 16802, U.S.A.}
\address{{2.} Physics Department, University of Utah, Salt Lake City, 
Utah 84112}
\address{{3.} Institute of Theoretical Physics, 
Warsaw University, ul. Ho\.{z}a 69, 00-681, Warsaw, Poland}
\address{{4.} Max Planck Institut  f\"ur Gravitationsphysik,
Am M\"uhlenberg,  D14476 Golm, Germany
} 
\maketitle

\begin{abstract}
    
Black hole mechanics was recently extended by replacing the more
commonly used event horizons in stationary space-times with isolated
horizons in more general space-times (which may admit radiation
arbitrarily close to black holes).  However, so far the detailed
analysis has been restricted to non-rotating black holes (although it
incorporated arbitrary distortion, as well as electromagnetic,
Yang-Mills and dilatonic charges).  We now fill this gap by first
introducing the notion of isolated horizon angular momentum and then
extending the first law to the rotating case.

\end{abstract}
\pacs{Pacs: 04070B, 0420}

\section{Introduction}
\label{s1}

The zeroth and first laws of black hole mechanics apply to equilibrium
situations and small departures therefrom.  In standard formulations
of these laws, black holes in equilibrium are represented by
stationary space-times with regular event horizons (see, e.g.,
\cite{review}).  While this idealization is a natural starting point,
from a physical perspective it seems quite restrictive.  (See
\cite{ack,abf} for a detailed discussion.)  To overcome this
limitation, a new model for a black hole in equilibrium was introduced
in \cite{ack,abf}.  The generalization is two-fold.  First, one
replaces the notion of an event horizon with that of an isolated
horizon.  While the former are defined only retroactively using the
fully evolved space-time geometry, the latter are defined
quasi-locally by suitably constraining the geometry of the horizon
surface itself.  Second, one drops the requirement that the space-time
be stationary and asks only that the horizon be isolated.  That is,
the requirement that the black hole be in equilibrium is incorporated
by demanding only that no matter or radiation fall through the horizon
although the exterior space-time region may well admit radiation.
Consequently, the generalization in the class of allowed space-times
is enormous.  In particular, space-times admitting isolated horizons
need not possess \textit{any} Killing vector field; although event
horizons of stationary black holes are isolated horizons, they are a
very special case.  A recent series of papers \cite{abf,afk} has
generalized the laws of black hole mechanics to this broader context.
The notion of isolated horizons has proved to be useful also in other
contexts ranging from numerical relativity to background independent
quantum gravity: i) it plays a key role in an ongoing program for
extracting physics from numerical simulations of black hole mergers
\cite{prl,abl1,abl3}; ii) it has led to the introduction
\cite{afk,acs,cs} of a physical model of hairy black holes, systematizing a
large body of results on properties of these black holes which has
accumulated from a mixture of analytical and numerical investigations;
and, iii) it serves as a point of departure for statistical mechanical
entropy calculations in which all non-rotating black holes (extremal
or not) \textit{and} cosmological horizons are incorporated in a
single stroke \cite{ack,abck,abk}.

The first treatment of black hole mechanics using isolated horizons
\cite{abf} only considered undistorted, non-rotating horizons.  That
is, the boundary conditions it used imply the \textit{intrinsic}
geometry of the horizon is spherically symmetric and the imaginary
part of the Weyl curvature component $\Psi_2$---which encodes
gravitational angular momentum---vanishes at the horizon.  Although
they do not constrain fields in the exterior region in any way (even
close to the horizon), these restrictions are, nonetheless, very
strong.  More recently, these boundary conditions were significantly
weakened to allow for both distortion and rotation and the basic,
geometric consequences of the more general boundary conditions were
analyzed \cite{afk}.  The zeroth law of black hole mechanics was also
extended to the more general context, and the first law, to
non-rotating but possibly distorted isolated horizons.  The purpose of
this paper is to extend the first law to the rotating case.  We will
first introduce (quasi-local) definitions of angular momentum and mass
of the isolated horizon in this context and then establish the first
law.  Thus, this paper is a continuation of \cite{afk} and completes
the task of deriving the black hole mechanics of all isolated horizons
of direct physical interest.

Let us outline the key new points which distinguish the rotating case
considered here.  First, if the (gravitational contribution to the)
horizon angular momentum is to be non-zero, the Weyl component
$\ImPt{\Psi_2}$ cannot vanish on the horizon.  Therefore, we will
extend the Hamiltonian framework of \cite{afk} by lifting the
restriction on $\ImPt{\Psi_2}$.  Second, by analogy with Killing
horizons, in the non-rotating case it was natural to require the
time-evolution vector field to lie along the null normal to the
horizon.  In the general rotating case, by contrast, we expect the
natural ``time-translation-like'' vector field to be space-like on the
horizon, with components both along the null generators
\textit{and} along a transverse ``rotational'' direction.  Our
evolution field will have this feature.  Third, to have a well-defined
notion of angular momentum, one expects there should exist a
rotational symmetry on the horizon.  We will analyze symmetries of all
isolated horizons and show that the boundary conditions imply that
there are three universal classes, characterized by the structure of
the symmetry group: I) spherically symmetric horizons; II)
axi-symmetric horizons; and, III) general, distorted horizons (with no
symmetry other than that along the null generators).  To have a
well-defined notion of angular momentum, we will focus on class
II. (Class I was discussed in \cite{abf} while the focus of
\cite{afk} was on \textit{non-rotating} isolated horizons in class III.)

In spite of these differences, the basic techniques used in this paper
are very similar to those of \cite{afk}.  The Hamiltonian formalism is
again employed to motivate the definition of horizon mass and, in our
rotating case, also angular momentum.  The first law again emerges as
a necessary and sufficient condition for the evolution to be
Hamiltonian, but now acquires new terms resulting from the angular
momentum of the horizon.

The plan of this paper is as follows.  Section \ref{s2} briefly
recalls the definition and basic structure of isolated horizons as
discussed in \cite{afk}.  Section \ref{s3} analyzes the possible
symmetries of isolated horizons.  Section \ref{s4} proves the first
law of black hole mechanics for space-times containing rotating
isolated horizons. Section \ref{s5} summarizes the results and the
Appendix discusses the issue of defining Hamiltonians generating
diffeomorphisms which need not be horizon symmetries.

\section{Preliminaries}
\label{s2}

This section summarizes the basic properties of isolated horizons and
introduces the notation used in this paper.  Specifically, subsection
\ref{s2.1} recalls from \cite{afk} the definition of a \textit{weakly
isolated horizon} in Einstein--Maxwell theory and several of its
immediate consequences.  Subsection \ref{s2.2} analyzes certain
geometric structures on the horizon.  This discussion is essential to
the classification of horizon symmetries presented in section \ref{s3}
and has not appeared before.  Finally, subsection \ref{s2.3} reviews
the covariant phase space of space-times admitting weakly isolated
horizons as inner boundaries, constructed in \cite{afk}.

Let us begin fixing a few conventions.  Throughout this paper, we
assume all manifolds and fields are smooth and restrict ourselves to
the Einstein--Maxwell theory.  Space-time $\M$ is a 4-dimensional
manifold equipped with a metric $g_{ab}$ of signature $(-, +, +, +)$
and a Maxwell potential $\emA_a$.  If $\D$ is a null hypersurface in
$(\M, g_{ab})$, its \textit{future-directed} null normal direction
will be denoted $\{\ell^a\}$.  This direction field is naturally an
equivalence class of vector fields on $\D$ under rescaling by
arbitrary positive functions.  We refer to it simply as the
\textit{null normal} to $\D$.  Pull-backs to $\D$ of the covariant
tensor fields on $\M$ (or more generally, covariant indices of
arbitrary tensor fields) will be denoted by an under-arrow and
equalities restricted to $\D$, by the symbol $\eqhat$.  The
(degenerate) intrinsic geometry on $\D$ is described by the metric
$q_{ab} :\eqhat g_{\pback{ab}}$.  A tensor $q^{ab}$ on $\D$ will be
said to be an ``inverse'' of $q_{ab}$ if it satisfies $q^{ab} q_{ac}
q_{bd} \eqhat q_{cd}$.  Because of the degeneracy of the intrinsic
geometry on $\D$, the inverse metric is not unique, but can be changed
freely by the addition of a term of the form $V^{(a} \ell^{b)}$ with
$V^a$ tangent to $\D$ and $\ell^b \in \{\ell^b\}$.  The expansion
$\theta_{(\ell)}$ of the null normal field $\ell^a \in \{\ell^a\}$ is
defined by $\theta_{(\ell)} :\eqhat q^{ab} \nabla_a \ell_b$, where
$\nabla_a$ is the torsion-free connection on $\M$ defined by $g_{ab}$.
It is straightforward to check this definition is independent of the
choice of inverse metric, but does depend upon the choice of null
normal vector.

\subsection{Weakly Isolated Horizons}
\label{s2.1}

A \textit{weakly isolated horizon} consists of a pair $(\D,
[\ell^a])$, where $\D$ is a 3-dimensional sub-manifold of $\M$ and
$[\ell^a]$ is an equivalence class of vector fields on $\D$ under
\textit{constant} rescalings, such that

\begin{enumerate}
  \renewcommand{\theenumi}{\roman{enumi}}
  \renewcommand{\labelenumi}{(\theenumi) }
  \renewcommand{\theenumii}{\alph{enumii}}
  \renewcommand{\labelenumii}{\theenumi\theenumii.\ }

\item\label{bcNul} $\D$ is topologically $S^2 \times \Re$ and null, 
$[\ell^a]$ lies along its null normal and the space of its integral
curves is diffeomorphic to $S^2$;

\item\label{bcExp} The expansion $\theta_{(\ell)}$ of $\D$ vanishes
for any choice $\ell^a \in [\ell^a]$ of the null normal;

\item\label{bcCon} The space-time connection has a (partial) symmetry
along $[\ell^a]$ in the sense that
\begin{equation}\label{conSym}
	[\Lie_\ell, \nabla_{\pback{a}}] \ell^b \eqhat 0
\end{equation}
for any choice of $\ell^a \in [\ell^a]$; 

\item\label{bcEoM} All equations of motion hold \textit{at} $\D$; and,

\item\label{bcMax} The Maxwell potential $\emA$ is
\textit{gauge-adapted} to the horizon in the sense that
\begin{equation}\label{maxSym}
	\Lie_\ell \emA_{\pback{a}} = 0
\end{equation}
for any choice of $\ell^a \in [\ell^a]$.

\end{enumerate}

A Killing horizon (with topology $S^2 \times \Re$) in
Einstein--Maxwell theory is automatically a weakly isolated horizon
under this definition, provided the Maxwell field strength is
symmetric along the Killing field and the gauge of its potential is
chosen to satisfy \eqref{maxSym}. If the Killing field is defined only
in a neighborhood of the horizon, there is no obvious way to fix the
freedom of rescaling it by a constant. This freedom is reflected in
our freedom to rescale $\ell$ by a constant. Note, however, the
definition admits a much broader class of examples.  To explore this
class, let us begin by examining the motivations behind the conditions
themselves.

Some of the restrictions made by the above conditions are relatively
tame.  For example, the topological requirement in condition
(\ref{bcNul}) simply restricts the horizon to have the topology which
one expects to arise from gravitational collapse.  This restriction
can be weakened to allow for more general topologies \cite{afk,kw},
though we shall not discuss this possibility further here.  More
importantly, the remainder of condition (\ref{bcNul}) makes clear the
roles of $\D$ and $[\ell^a]$: $\D$ is the horizon surface and
$[\ell^a]$ is a distinguished class of its null normals.  Condition
(\ref{bcEoM}) is also straightforward; it applies a dynamical
restriction closely analogous to the one usually imposed at infinity.
However, while the metric at infinity is required to approach a
\textit{specific} asymptotic solution to the Einstein equations,
condition (\ref{bcEoM}) allows the metric to approach \textit{any}
solution to the Einstein equations at the horizon.

Conditions (\ref{bcNul}) and (\ref{bcEoM}) are satisfied on a wide
variety of surfaces, including many even in Minkowski space-time.
However, the vast majority of these surfaces do not have the
characteristics one would intuitively expect on a ``horizon.''  The
first key condition which distinguishes a weakly isolated horizon is
(\ref{bcExp}).  It implies the cross-sectional area of $\D$ is
\textit{constant} ``in time,'' thereby capturing the notion of
isolation without introducing a Killing field.  The horizon area is
denoted $\AD$ and we define its radius $\RD$ by
$$\AD = 4\pi\RD^2.$$ 
This condition makes the definition dramatically stronger.  For
example, it implies there are in fact no weakly isolated horizons in
Minkowski space-time.

To discuss condition (\ref{bcCon}), it will be useful to first explore
some consequences of condition (\ref{bcExp}) by itself.  First, since
$\ell^a \in [\ell^a]$ is normal to $\D$, it is automatically
twist-free and geodetic:
\begin{equation}\label{sgDef}
	\ell^a \nabla_a \ell^b \eqhat \kappa_{(\ell)} \ell^b.
\end{equation}
Motivated by the usual definition for Killing horizons, the
acceleration $\kappa_{(\ell)}$ of $\ell^a$ will be called the
\textit{surface gravity of} $\D$ \textit{with respect to the normal}
$\ell^a$.  As one can see from \eqref{sgDef}, when $\ell^a$ is
rescaled within $[\ell^a]$, the surface gravity scales by the same
factor.  Thus, the surface gravity of a weakly isolated horizon is
generally defined only up to a positive constant multiplicative
factor.  Second, given the vanishing twist and expansion of $\ell^a$,
one can use the Raychaudhuri equation for the null congruence
generated by $\ell^a$ to conclude its \textit{shear} must also vanish,
and that
\begin{equation}\label{RicNul}
	R_{ab} \ell^a \ell^b \eqhat 0.
\end{equation}
Third, the vanishing twist, shear and expansion of $\ell^a$ imply the
existence of a 1-form $\omega_a$, intrinsic to $\D$, satisfying
\begin{equation}\label{hcDef}
	\nabla_{\pback{a}} \ell^b \eqhat \omega_a \ell^b.
\end{equation}
This 1-form $\omega_a$ is independent of the choice of $\ell^b \in
[\ell^b]$ and its contraction with $\ell^a$ will yield the surface
gravity $\kappa_{(\ell)}$ defined above.  Furthermore, as we will now
discuss, condition (\ref{bcCon}) causes $\omega_a$ to play a central
role in the theory of weakly isolated horizons.

One important consequence of \eqref{hcDef} is that $\ell^a$ is a
symmetry of the degenerate intrinsic geometry of $\D$ in the sense
that
\begin{equation}\label{qSym}
	\Lie_\ell q_{ab} \eqhat 0.
\end{equation}
In the case of a non-null hypersurface in space-time on which the
intrinsic geometry is non-degenerate, the existence of a Killing
vector for the intrinsic metric would automatically imply the
intrinsic connection was likewise preserved along that Killing vector.
In other words, for a non-degenerate intrinsic geometry, the
restriction \eqref{conSym} of condition (\ref{bcCon}) would
\textit{follow} from \eqref{qSym}.  However, this is not the case for
a null hypersurface since the intrinsic metric does not determine the
intrinsic connection uniquely.  Thus, condition (\ref{bcCon}) must be
imposed separately.  The logic of making this restriction, however, is
clear: we are extending the symmetry \eqref{qSym} of the degenerate
metric to the intrinsic connection on $\D$ given by
$\nabla_{\pback{a}}$.  Note, however, that condition (\ref{bcCon}) does
not restrict the \textit{entire} connection on $\D$, but only its
action on $\ell^b$.  In fact, given the definition \eqref{hcDef},
condition (\ref{bcCon}) can be written simply as
\begin{equation}\label{hcSym}
	\Lie_\ell \hc_a = 0.
\end{equation}
This formula is much easier to apply in practice than \eqref{conSym},
which is why the 1-form $\hc_a$ plays an important role in our
formulation.  (If one requires that $\ell^a$ be a symmetry of the full
pull-back of the connection to $\D$, we obtain isolated horizons
\cite{afk,abl1}.  While this stronger condition is physically
reasonable, it is significantly more difficult to check its validity
in examples.  Since the stronger condition is not needed in our proof
of the zeroth and the first laws, we have chosen to work with weakly
isolated horizons.)

Finally, let us examine condition (\ref{bcMax}) on the Maxwell field.
At first, the symmetry requirement \eqref{maxSym} on the Maxwell field
appears similar to the restriction \eqref{hcSym} placed on the horizon
geometry by condition (\ref{bcCon}).  However, the two are actually
quite different.  While \eqref{hcSym} represents a genuine restriction
on the physical fields at the horizon, \eqref{maxSym} can always be
achieved via a gauge transformation when the other conditions are
satisfied.  Using the Einstein equations at the horizon and the
consequence \eqref{RicNul} of the Raychaudhuri equation one can show
the 1-form $\ell^a \emF_{ab}$ is null at the horizon.  Since this
1-form is also orthogonal to $\ell^b$, it must be proportional to
$\ell_b$, whence\relax
\footnote{The symbol $\cdot\vins\cdot$ represents the contraction of a
vector field on the first index of a differential form.}
\begin{equation}\label{pbF}
	\ell \vins \pback{\emF} \eqhat 0.
\end{equation}
Then, using the Cartan formula for the Lie derivative and the Maxwell
equations at the horizon, one finds
$$	\Lie_\ell \pback{F} \eqhat \ell \vins \pback{\ed\emF} + 
		\pback{\ed (\ell \vins \emF)} = 0. $$
Thus, because the field strength is already compatible with the
symmetry \eqref{maxSym} imposed on the Maxwell potential by condition
(\ref{bcMax}), \eqref{maxSym} is indeed merely a (partial) choice of
gauge.

We conclude this subsection with several remarks.  Each of these
points is discussed in more detail in \cite{afk}.

First, while this paper explicitly examines only those weakly isolated
horizons which occur in Einstein--Maxwell theory, it is not difficult
to generalize the definition given here to allow other matter fields
near the horizon.  However, it is not possible to incorporate
arbitrary types of matter, but only those which satisfy a certain
energy condition.  Specifically, the stress-energy tensor of each
matter field present at the horizon must have the property that
$T_{ab} \ell^b$ is future directed and causal\relax
\footnote{A vector is said to be \textit{causal} if it is either
time-like or null.}
for any $\ell^b \in [\ell^b]$.  This condition follows immediately
from the dominant energy condition which is usually assumed in proofs
of the laws of black hole mechanics.  Moreover, since the Maxwell (or
Yang-Mills, etc.)  field satisfies the dominant energy condition, it
will automatically meet the (weaker) demand made here.  This condition
is essential to the derivation of \eqref{RicNul} from the Raychaudhuri
equation and to the associated proof that the horizon is shear-free.

Second, the conditions given above imply the \textit{space-time}
geometry at a weakly isolated horizon is algebraically special.  This
is an interesting feature which carries over from the stationary
context where the Kerr--Newman black hole space-times are (globally)
of Petrov type II-II. However, there are several important
differences.  One can show $[\ell^a]$ is a repeated principal null
direction \textit{at the horizon}, whence the space-time geometry is
necessarily of Petrov type II there.  However, it is not generally
possible to identify a second repeated principal null direction at
such a horizon, so the geometry needn't be of type II-II. Moreover,
the space-time geometry \textit{away} from the horizon may not be
algebraically special at all; our conditions constrain only the local,
horizon geometry.  Since $[\ell^a]$ is a repeated principal null
direction for the space-time metric at the horizon, the
Newman--Penrose curvature components satisfy
$$	\Psi_0 \eqhat \Psi_1 \eqhat 0 $$
in any null frame which includes an element $\ell^a \in [\ell^a]$ at
$\D$.  It then follows that the Newman--Penrose component $\Psi_2$ is
``gauge-invariant'' (i.e., independent of the other three null tetrad
elements) at $\D$.  In particular, the imaginary part of $\Psi_2$ can
be expressed without reference to any null tetrad using the 1-form
$\hc_a$ introduced in \eqref{hcDef}:
\begin{equation}\label{dhc}
	\ed\hc = 2\ImPt{\Psi_2} \te,
\end{equation}
where $\te_{ab}$ denotes the natural area 2-form on $\D$ (see the next
subsection).

Third, the conditions enable a simple proof of the zeroth law of black
hole mechanics for generic weakly isolated horizons.  Examining the
definition \eqref{sgDef} of surface gravity for a weakly isolated
horizon and the definition \eqref{hcDef} of $\hc_a$, one finds
\begin{equation}\label{hcsg}
	 \kappa_{(\ell)} = \ell^a \hc_a \, ,
\end{equation}
whence (by the Cartan formula) we have
$$	\ed \kappa_{(\ell)} \equiv \ed(\ell \vins \hc) = 
\Lie_\ell \hc - \ell \vins \ed\hc. $$
The first term on the right hand side vanishes because of
\eqref{hcSym}, while the second vanishes because $\ed\hc$ is given by
\eqref{dhc} and the contraction of $\te_{ab}$ with $[\ell^a]$
vanishes.  Thus, the surface gravity of a weakly isolated horizon is
indeed constant; the zeroth law holds.  Note, however, that because
$\D$ is equipped only with an \textit{equivalence class} $[\ell^a]$ of
null normals, we have the constant rescaling freedom $\ell^a
\rightarrow \tilde\ell^a = c \ell^a$ under which the surface gravity
transforms via $\kappa_{(\ell)} \rightarrow \kappa_{(\tilde\ell)} = c
\kappa_{(\ell)}$.  Therefore, while surface gravity is constant on an
isolated horizon, we can not assign a particular value to it.  This is
not surprising: one cannot assign a specific numerical value to
surface gravity even on a \textit{locally} defined Killing horizon,
since the Killing field is defined only up to rescaling by a constant.
Finally, we note the argument establishing the zeroth law for weakly
isolated horizons can also be applied to the Maxwell field.  Repeating
this argument, with $\hc$ replaced by $\emA$, and applying
\eqref{maxSym} and \eqref{pbF} one finds the quantity $\ell \vins
\emA$ is also constant over the horizon surface.  Motivated by this
fact, we define the \textit{electric potential} of the horizon by
\begin{equation}\label{epDef}
	\Phi_{(\ell)} := -\ell^a \emA_a.
\end{equation}
Once again, while the electric potential is always constant over $\D$, its
value is not fixed.

\subsection{Geometrical Structures} 
\label{s2.2}

We will first discuss the relation between certain types of fields on
$\Delta$ and fields on the 2-sphere $\PD$ of integral curves of
$[\ell^a]$ and then show that if $\kappa_{(\ell)}$ is non-zero, $\D$
admits a natural foliation.  This discussion will be useful especially
in section \ref{s3} for our analysis of the symmetry algebras of
weakly isolated horizons.

Denote by $\gen$ the natural projection map $\gen$ from $\D$ to $\PD$.
A vector field $W^{a}$ defined intrinsically on $\D$ can be
unambiguously projected to a vector field $\PDp W^{a} $ on $\PD$ if
and only if $\Lie_\ell {W^{a}}$ is proportional to $\ell^a$.
Similarly, a covariant tensor field $T_{a_1\cdots a_n}$ on $\D$ is a
pull-back under $\gen$ of a tensor field $\PDp T_{a_1 \cdots a_m}$ on
$\PD$ if and only if two conditions hold: i) $T_{a_1 \cdots a_m}{}$ is
transversal to $\ell$ in the sense that the contraction of any of its
indices with $\ell$ vanishes; and, ii) the Lie derivative $\Lie_\ell
{T_{a_1 \cdots a_m}}$ vanishes.  These are rather general properties
of manifolds ruled by one-dimensional curves.

Applying them to the case of a weakly isolated horizon, we conclude
that the degenerate metric $q_{ab}$ is the pull-back to $\D$ of a
Riemannian metric $\hat{q}_{ab}$ on $\PD$.  The connection 1-form
$\hc_a$ does satisfy condition ii) above.  However, since
$$	\ell^a \hc_a \eqhat \kappa_{(\ell)}, $$
it defines a 1-form $\PDp\hc_a$ on $\PD$ \textit{only} in the extremal
case, i.e., only when $\kappa_{(\ell)} \eqhat 0$.  Finally, the
projection map $\gen$ enables us to define a preferred area element
$\te_{ab}$ on $\D$.  Using the Riemannian metric $\PDp q_{ab}$, one
can construct the area element $\hat\epsilon_{ab}$ on $\PD$ and pull
that 2-form back to $\D$ under $\gen$.  By construction, the resulting
2-form $\te_{ab}$ on $\D$ satisfies
$$ \ell^a \te_{ab} = 0 \qquad\mbox{and}\qquad 
\Lie_\ell \te_{ab} = 0. $$
The first of these results was already used in the proof of the zeroth
law in the last subsection.

Let us now assume that $\kappa_{(\ell)}$ is non-zero and discuss the
preferred foliation of $\Delta$ (for further details and a discussion
of global subtleties, see \cite{abl1}).  For simplicity, consider
first the ``non-rotating case'' where $\ImPt{\Psi_2} = 0$.  According
to \eqref{dhc}, this is precisely the case where the connection form
$\hc_a$ is curl-free.  Hence $\hc = \ed v$ for some function $v$ on
$\D$.  Furthermore, $\ell \vins \ed v = \kappa_{(\ell)}$ is a non-zero
constant on $\D$, whence $[\ell^a]$ is transverse to the $v = {\rm
const}$ surfaces.  Thus, $v = {\rm const}$ surfaces define a natural
foliation of $\D$ by a family of 2-spheres.

To conclude, let us consider the rotating case.  While $\ImPt{\Psi_2}$
no longer vanishes in this case, it is still constant along each
generator of $\D$, i.e., is a pull-back of a function $\PDp
{\ImPt{\Psi_2}}$ on $\PD$.  Using (\ref{dhc}) as the motivation, let
us introduce a 1-form $\PDp \alpha_a$ on $\PD$ such that $\PDp\ed
\PDp\alpha = 2 \PDp {\ImPt{\Psi_2}}\, \PDp\epsilon$.  Of course, this
potential $\PDp \alpha_a$ is not unique; one is free to add to it the
gradient of any function on $\PD$.  There is, however, a natural gauge
condition which fixes this ambiguity: Since there are no harmonic
1-forms on the 2-sphere $\PD$, there exists a unique 1-form
$\hat{\alpha}_a$ such that
\begin{equation}\label{dfDef}
\PDp{\ed} \hat{\alpha} = 2\PDp{\ImPt{\Psi_2}} \PDp\epsilon 
\qquad\mbox{and}\qquad  \PDp{\ed} \PDp\dual\hat{\alpha} = 0,
\end{equation}
where $\PDp\epsilon_{ab}$ is the area element on $\PD$ and $\PDp\dual$
denotes the Hodge dual on $\PD$.  Let $\alpha_a$ be the pull-back to
$\D$ of $\hat\alpha_a$ and set
%
$$ \tilde{\hc}_a = \hc_a - \alpha_a. $$
%
Then, $\ed \tilde{\hc}_a = 0$ and $\ell^a \tilde\hc_a =
\kappa_{(\ell)}$ on $\D$.  Thus, $\tilde{\hc}$ has the same properties
that $\hc_a$ had in the non-rotating case.  Therefore, it defines a
foliation of $\D$ by a family of 2-spheres.  Our gauge choice is
natural in the sense that it is the only choice for which this more
general construction, when applied to the non-rotating case,
reproduces the natural foliation obtained above.  (Indeed, in the
non-rotating case and in our gauge, $\hat{\alpha}_a$ vanishes on $\PD$
and $\tilde\hc_a = \hc_a$ on $\D$.)  While this foliation is used only
to elucidate certain points in this paper, it plays a key role in the
analysis of the near-horizon, strong field geometry \cite{abl3} and in
extracting physics from space-times obtained via numerical
simulations, such as those associated with black hole mergers
\cite{abl3,prl}.

\subsection{Covariant Phase Space}
\label{s2.3}

In the spirit of \cite{afk}, we will use Hamiltonian methods to
introduce notions of angular momentum and mass of weakly isolated
horizons.  In this sub-section, we recall from \cite{afk} the
structure of the underlying phase space.

\begin{figure} \label{f1}
  \begin{center}
  \includegraphics[height=4cm]{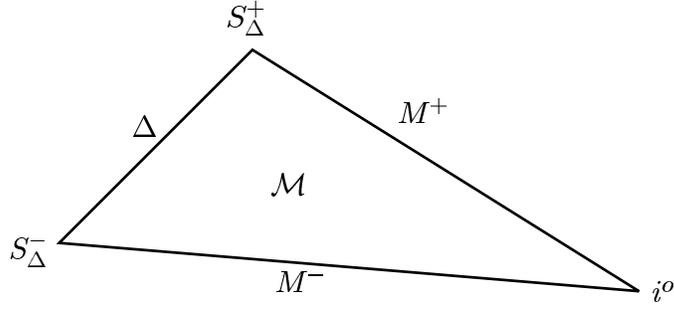}
	\bigskip
  \caption{The region of space-time $\M$ under consideration has an
  internal boundary $\Delta$ and is bounded by two partial Cauchy
  surfaces $M^{\pm}$ which intersect $\Delta$ in the $2$-spheres
  $S^{\pm}_\D$ and extend to spatial infinity $i^{o}$. }
  \end{center}
\end{figure}

The most interesting space-times containing weakly isolated horizons
are of the black hole type, so we will tailor the topological
structure of the underlying space-time manifold to this case.
Specifically, fix a manifold $\tilde{\M}$ with boundary where the
boundary consists of four components: an internal boundary
$\tilde{\D}$, topologically $S^2 \times \Re$ which will finally serve
as the isolated horizon, an outer boundary $\tau_\infty$, also
topologically $S^2\times \Re$ which will serve as a time-like cylinder
at spatial infinity, and two surfaces $M^{\pm}$, which are intended to
be space-like and to serve as future and past boundaries of $\M$.  The
intersections of $M^\pm$ with $\D$ and $\tau_\infty$ will be denoted
$\SD^\pm$ and $\Si^\pm$, respectively (see FIG 1).  The inner boundary
$\D$ will be equipped with a privileged equivalence class $[\ell^a]$
of vector fields under constant rescalings, whose integral curves
constitute a 2-sphere $\PD$.  In the Hamiltonian framework, we will
not be interested in the surfaces $M^\pm$ and $\tau_\infty$.
Therefore, it is convenient to introduce $\M := \tilde{\M} - (M^+ \cup
M^- \cup \tau_\infty)$ and $\D :=
\tilde{\D} - (S^+_\D \cup S^{-}_\D)$.

We will consider Einstein--Maxwell theory in a first-order framework
based on tetrads.  Therefore, our basic physical fields will consist
of a triple $(e^a_I, A_a{}_I{}^J, \emA_a)$ on $M$, where $e^a_I$ is a
tetrad on space-time, $A_a{}_I{}^J$ is the connection form (relative
to a fiducial flat connection $\partial_a$) for a connection $D_a$ in
the frame bundle over space-time, and $\emA_a$ is a Maxwell connection
on a trivial $U(1)$ bundle over $\M$ (see \cite{afk}).  (Whenever
needed, these fields will be extended to the boundaries $S_{\D}^\pm$
of $\D$ by continuity.)  These fields are subject to certain boundary
conditions: We require that each space-time $(\M, e^a_I)$ is
asymptotically flat at spatial infinity and admits $(\D, [\ell^a])$ as
a weakly isolated horizon.  For convenience, we will fix an
\textit{internal} tetrad $(\ell^I, n^I, m^I, \bar m^I)$ at each point 
of $\D$ and require the tetrad field to satisfy $\ell^I e^a_I \in
[\ell^a]$ on $\D$.  (This can always be achieved and serves to
eliminate an irrelevant part of the tetrad-rotation freedom.)

With this kinematical structure at hand, we can now specify the action
functional:
\begin{equation}\label{act}
S[e, D, \emA]\, = \, \frac{-1}{16\pi G} \int_\M \Tr{\Sigma \wedge F} + 
	\frac{1}{16\pi G} \int_{\tau_\infty} \Tr{A \wedge \Sigma} - 
	\frac{1}{8\pi} \int_\M \emF \wedge \dual\emF,
\end{equation}
where the traces are taken over the internal (tetrad) indices,
$F_{ab}{}_I{}^J$ denotes the curvature of the connection
$A_a{}_I{}^J$, and the 2-forms $\Sigma_{ab}{}_I{}^J$ are defined in
terms of the (co-)tetrad field by
%
$$\Sigma_{ab}{}_I{}^J := \epsilon_I{}^J{}_{KL} e_a^K e_b^L. $$
%
Interestingly, while the asymptotic boundary term is needed to make
the action principle well-posed given the asymptotically flat fall-off
conditions imposed near infinity, no such boundary term is needed at
the horizon surface $\D$.  (For details, see \cite{afk}.)

Our covariant phase space $\G$ will consist of the space of solutions
$(e^a_I, A_a{}_I{}^J, \emA_a)$ to the Einstein-Maxwell equations,
satisfying the boundary conditions specified above.  As usual, the
symplectic structure is constructed on $\G$ using the
(anti-symmetrized) second variation of the action \eqref{act}.
Applying the equations of motion to this second variation, one
discovers the integral over $\M$ reduces to surface terms at $M^\pm$
and at $\D$ (the surface term at $\tau_\infty$ vanishes because of the
asymptotic fall-off conditions).  In order to reduce the expression
for the second variation further, it is convenient to introduce a pair
of scalar functions on $\D$ which act as ``potentials'' for the
surface gravity and electric potential.  Given any point $(e^a_I,
A_a{}_I{}^J, \emA_a)$ in phase space, the scalar field $\psi$ is
defined by
\begin{equation}\label{grgDef}
\psi = 0 \quad\mbox{on $\SD^-$, and}\quad \Lie_\ell \psi = 
\kappa_{(\ell)}
\end{equation}
and the scalar field $\chi$ by 
\begin{equation}\label{emgDef}
\chi = 0 \quad\mbox{on $\SD^-$, and}\quad \Lie_\ell \chi = -\Phi_{(\ell)}.
\end{equation}
Note that both of these functions are \textit{completely determined by
the physical fields}. (A priori there is some freedom in the choice of
initial values of $\psi$ and $\chi$ on $S^{-}_\Delta$. While the
symplectic structure is sensitive to these choices, none of the final
results are.)  Expressed in terms of $\psi$ and $\chi$, the surface
term at $\D$ in the second variation of the action turns out to be
exact and thus reduces to a pair of integrals on $\SD^\pm$.  The
integral over $M^-$, together with its surface term at $\SD^-$ is then
taken to define the symplectic structure for the theory.  In fact,
when the equations of motion hold and both variations satisfy the
linearized equations of motion, the integral which defines the
symplectic structure may be taken over any partial Cauchy surface $M$
in $\M$.  It is given by
\begin{eqnarray}\nonumber
\Omega(\delta_1, \delta_2) &=& \frac{-1}{16\pi G}\, \int_M 
\Tr{\delta_1 A \wedge \delta_2 \Sigma - \delta_2 A \wedge \delta_1 \Sigma}
- \frac{1}{8\pi G}\, \oint_{\SD} 
\delta_1 \psi \delta_2 \te - \delta_2 \psi \delta_1 \te 
\\&&\label{symp}{}
+ \frac{1}{4\pi}\, \int_M 
\delta_1\emA \wedge \delta_2(\dual\emF)  - 
\delta_2 \emA \wedge \delta_1(\dual\emF)
- \frac{1}{4\pi}\, \oint_{\SD} 
\delta_1\chi \delta_2(\dual\emF) - \delta_2\chi \delta_1(\dual\emF),
\end{eqnarray}
where $M$ intersects $\D$ in $\SD$.  Details of this result are
discussed in \cite{afk}.

In the covariant Hamiltonian framework, infinitesimal gauge
transformations are in 1-1 correspondence with the degenerate
directions of the symplectic structure. For example in the
asymptotically flat context \textit{without} internal boundaries,
infinitesimal diffeomorphisms and internal Lorentz rotations are in
the kernel of the symplectic structure if and only if they vanish
asymptotically. Thus, infinitesimal diffeomorphisms which
asymptotically approach Killing fields of the flat background metric
are not gauge; they are generated by non-trivial Hamiltonians which
define the ADM 4-momentum and angular momentum. The situation is
similar at the internal boundary $\D$. Infinitesimal diffeomorphisms
and internal Lorentz rotations need not be in the kernel of the
symplectic structure unless they vanish on $\D$. Indeed, we will show
in Section \ref{s4} that space-time motions corresponding to time
translations and rotations are generated by non-trivial Hamiltonians
with surface terms both at infinity \textit{and} at the horizon. The
horizon surface terms will then be interpreted as the horizon energy
and angular momentum. However, since infinitesimal internal rotations
(and gauge transformations of the Maxwell theory) on $\D$ are not
automatically in the kernel of the symplectic structure, one must
separately ensure that physical quantities such as energy and angular
momentum are gauge invariant.

\section{Symmetries of a Weakly Isolated Horizon}
\label{s3}

In this section we analyze the possible symmetries of a weakly
isolated horizon.  Because the generators $[\ell^a]$ of $\D$ are not
assumed to be complete, we will focus on \textit{infinitesimal}
symmetries.  Since the horizon is the inner boundary of space-times
under consideration while spatial infinity is the outer, one would
expect the horizon symmetries to be analogous to the asymptotic
symmetries at spatial infinity.  Thus, it is natural to define the
horizon symmetry algebra ${\L}$ to be the quotient of the Lie algebra
of all infinitesimal space-time diffeomorphisms which preserve the
horizon structure by its sub-algebra consisting of elements which
vanish on the horizon.  The key question now is: What is the horizon
structure relevant for these considerations?  It is here that one
would expect a significant departure from the situation at spatial
infinity.  At infinity, all metrics approach a fixed flat metric
whence the relevant geometric structure ---and hence also the symmetry
algebra--- is universal; it does not vary from one space-time to
another.  At the horizon, by contrast, we are in the strong field
regime and the near-horizon geometry can vary from one space-time to
another.  Therefore, we do not expect the relevant horizon structure
or the symmetry algebra ${\L}$ to be universal.  Nonetheless, because
the horizon \textit{is} a boundary and all geometries under
consideration are subject to the same boundary conditions, one would
expect the horizon symmetry algebra to fall in a small number of
``universality classes''.  We will see that this expectation is
correct.

To identify the relevant horizon structure, let us return to the
definition of weakly isolated horizons and examine the geometric
structures that are essential to the definition.  First, we have the
manifold $\D$ and the equivalence class $[\ell^a]$ of null normals.
This structure is shared by all weakly isolated horizons.  However, to
specify the conditions that $[\ell^a]$ must satisfy for $(\D,
[\ell^a])$ to qualify as an isolated horizon, we also introduced
``non-universal'' fields $(q_{ab}, \hc_a)$ which vary from one point
in our covariant phase space to another.
\footnote{Recall from Section \ref{s2.1} that the essential
restrictions are captured by two conditions: i) $\Lie_\ell q_{ab}
\eqhat 0$; and, ii) $\Lie_\ell \omega_a \eqhat 0$.}
Therefore, to construct the algebra ${\L}$ corresponding a given
isolated horizon, it is natural to focus on space-time vector fields
whose action preserves $(\D, [\ell^a], q_{ab}, \omega_a)$ and factor
out by those which vanish on $\D$.

Thus, we can ignore the space-time manifold $\M$ and work just with
vector fields $W^a$ which are tangential to $\D$.  To qualify as
symmetries, they should satisfy the following two conditions:
\begin{equation}\label{kinW}
\Lie_W \ell^a\, \eqhat c_W\, \ell^a \quad \mbox{with $c_W$ 
\textit{constant} on $\D$};
\end{equation}
and,
\begin{equation}\label{dynW}
\Lie_W q_{ab}\, \eqhat 0  \quad\mbox{and}\quad 
\Lie_W \, \hc_a \eqhat 0.
\end{equation}
The set of vector fields $W^a$ satisfying both \eqref{kinW} and
\eqref{dynW} form a Lie algebra under the usual commutator bracket.
This is the symmetry Lie algebra for the horizon in question.  We will
denote it by $\sa$ and refer to these vector fields $W^a$ as
infinitesimal symmetries of the horizon.

Before analyzing the full algebra, let us note that, irrespective of
the specific weakly isolated horizon under consideration, $\sa$ is at
least one-dimensional: By setting $W^a = \ell^a$ for any $\ell^a \in
[\ell^a]$, we see (\ref{kinW}) and (\ref{dynW}) are automatically
satisfied by virtue of the very definition of weakly isolated
horizons.  These infinitesimal symmetries preserve each integral curve
of $\ell$.  More generally, denote by $\sai$ all elements $W^a$ of
$\sa$ of the form $W^a = f \ell^a$ for some function $f$ on $\D$.
This will be the sub-algebra of all \textit{generator-preserving}
symmetries.  Now, given any infinitesimal symmetry $W^a$ in $\sa$ and
an infinitesimal symmetry $f\ell^a$ in $\sai$, (\ref{kinW}) implies
the commutator $[W, f\ell]^a$ is again proportional to $\ell^a$:
%
$$	[W, f\ell]^a \eqhat (c_W f + \Lie_W f) \ell^a, $$
%
where $c_W$ is the constant appearing in \eqref{kinW}.  Hence, $\sai$
is in fact an ideal of the Lie algebra $\sa$.  Denote by ${\psa}$ the
quotient Lie algebra: $\psa = \sa/\sai$.  For reasons explained below,
we will refer to it as the \textit{algebra of projected symmetries}.
Thus, the algebra $\sa$ of infinitesimal horizon symmetries is a
semi-direct sum\relax
\footnote{Had $\D$ been complete, we could have integrated $\L$ 
to obtain a horizon symmetry group ${\cal G}$ which would have been a
semi-direct \textit{product} of the groups generated by $\sai$ and
$\psa$, the former being the normal sub-group of ${\cal G}$.}
of $\sai$ and $\psa$. 

To determine the structure of $\sa$, therefore, it suffices to examine
the Lie algebras $\sai$ and $\psa$ separately.  Let us begin with the
former.  Setting $W^a = f\ell^a$ in (\ref{kinW}) we obtain:
\begin{equation} \label{ideal}
\Lie_\ell f = C_f 
\end{equation}
where $C_f$ is a constant on $\D$.  The first of the two conditions in
(\ref{dynW}) does not restrict $f$ in any way while the second
implies:
\begin{equation}\label{flhc}
0 \eqhat \Lie_{f\ell} \hc \eqhat \ed(f\ell \vins \hc) 
\eqhat \kappa_{(\ell)} \ed f\, .
\end{equation}
Therefore, if $\kappa_{(\ell)} \not= 0$, then $f$ must be constant on
$\D$.  In this case the only generator-preserving symmetries are given
by $W^a \in [\ell^a]$ and $\sai$ is 1-dimensional.  On the other hand,
if $\kappa_{(\ell)} = 0$ (i.e., $(\D, [\ell])$ is extremal),
(\ref{flhc}) imposes no condition on $f$.  In this case, $f$ has only
to satisfy \eqref{ideal} and $\sai$ is infinite-dimensional\relax

Next, let us examine the quotient Lie algebra $\psa$.  Since every
infinitesimal symmetry $W^a$ must satisfy \eqref{kinW}, it can be
projected unambiguously to a vector field $\PDp W^a$ on $\PD$.  We can
now use our discussion of Section \ref{s2.2} on the relation between
fields on $\D$ with those on the 2-sphere $\PD$ of its generators to
analyze $\psa$ in terms of these projections $\PDp W^a$.  The first of
Eqs (\ref{dynW}) is satisfied if and only if $\PDp W^a$ is a Killing
field on $(\PD, \PDp q_{ab})$:
%
$${\Lie_{\PDp W}\, \PDp{q}_{ab}} = 0\, $$
%
Denote by $\pka$ the isometry Lie algebra of $(\PD, \PDp q_{ab})$.  We
have shown that $\psa$ is a sub-algebra of $\pka$.  In general, it
will only be a proper sub-algebra because $W^a$ must also satisfy
$(\ref{dynW})$.  Nonetheless, the fact that $\psa$ is a sub-algebra of
$\pka$ plays a key role in the classification of $\sa$ because the
dimension and \textit{topology} of $\PD$ imposes severe restrictions
on $\pka$.  On the one hand, every Killing field $\PDp W^a$ on $\PD$
is of the form $\hat{\epsilon}^{ab} \partial_b \hat{h}$ for some
function $\hat{h}$.  On the other hand, since all metrics on a
2-sphere are conformal to one another, any Killing field for a generic
metric must be a conformal Killing field of a fixed round 2-sphere
metric on that sphere.  Every conformal Killing vector field
$\PDp{W}^a$ of a round metric belongs to one of the following classes:
(i) a rotation; (ii) a combination of a boost and rotation commuting
with each other, with a non zero boost component; and, (iii) a null
rotation, the set of zeros of the vector field has exactly one
element.  Now, in cases (ii) and (iii), \textit{all} the orbits of the
vector field converge to a same point of the sphere at which
$\PDp{W}^a=0$ (there are two such points in the case (ii)).  Since the
function $h$ is constant on every orbit, it follows that in these
cases it would be constant whence $\PDp{W}^a=0$ on the entire sphere.
Therefore, a non-trivial Killing vector field $\PDp{W}^a$ is of the
class (i).

Thus, we conclude that a general metric on a 2-sphere is of one of three
types:
\begin{enumerate}
	\item The metric is round; $\pka$ is three-dimensional and 
isomorphic to
	${\bf so}(3)$;

	\item The metric is axially symmetric; $\pka$ is one-dimensional 
and isomorphic to ${\bf so}(2)$;

	\item The metric has no Killing fields; $\pka$ is zero-dimensional 
(i.e., consists only of the zero element).
\end{enumerate}
In each case, the sub-algebras of $\pka$ are easy to characterize.
Only in the first case does $\pka$ admit a non-trivial sub-algebra
which then must be isomorphic to ${\bf so}(2)$.  Hence, the quotient
algebra $\psa$ must be isomorphic either to ${\bf so}(3)$ or ${\bf
so}(2)$ or must be the trivial Lie algebra consisting of only of the
zero element.

Thus we can divide the set of all weakly isolated horizons into the
three classes:
\begin{enumerate}
	\renewcommand{\theenumi}{\Roman{enumi}}
	\renewcommand{\labelenumi}{\theenumi{.} }
  \renewcommand{\theenumii}{\alph{enumii}}
  \renewcommand{\labelenumii}{\theenumi\theenumii.\ }

\item Spherically symmetric; the algebra $\psa$ of projected
symmetries is 3-dimensional. If $\kappa_{(\ell)} \not= 0$, the horizon
symmetry algebra ${\L}$ is the 4-dimensional semi-direct sum of the
additive group $\Re$ of reals with ${\bf so(3)}$. If $\kappa_{(\ell)}=
0$, then ${\L}$ is the semi-direct sum of the infinite dimensional Lie
algebra defined by \eqref{ideal} with ${\bf so(3)}$.

\item Axi-symmetric; $\psa$ is 1-dimensional. If $\kappa_{(\ell)}
\not= 0$, the horizon symmetry algebra ${\L}$ is the 2-dimensional
Abelian Lie algebra generated by vector fields $a\ell^a + b\phi^a$
where $\phi^a$ is a rotation on $\D$ and $a,b$ constants.  If
$\kappa_{(\ell)}= 0$, then ${\L}$ is the semi-direct sum of the
infinite dimensional Lie algebra defined by \eqref{ideal} with ${\bf
so(2)}$.

\item Generic; $\psa$ is zero-dimensional. If $\kappa_{(\ell)} \not=
0$, the horizon symmetry group ${\L}$ is 1-dimensional. If
$\kappa_{(\ell)}= 0$, then ${\L}$ is the infinite dimensional Lie
algebra defined by \eqref{ideal}
\end{enumerate}
\noindent As mentioned in the Introduction, in the next section, we
will focus on class II.

We will conclude with two remarks.

1.  In the above discussion, we have focussed only on symmetries of
the relevant horizon geometry.  In the Einstein--Maxwell theory, it is
natural to require that an infinitesimal symmetry $W^a$ of a given
weakly isolated horizon also satisfy $\Lie_W F_{\pback{ab}} \eqhat 0$.
Physically, this condition requires the symmetry to preserve the flux
of the magnetic field through the horizon.  It is natural to demand
the same of the electric flux.  Therefore, we will also insist that
$\Lie_W (\dual F)_{\pback{ab}} \eqhat 0$ at the horizon.  Neither of
these additional conditions will affect the classification scheme
described above in any way.

2. As shown in Section \ref{s2.2}, if $\kappa_{(\ell)} \not=0$, the
weakly isolated horizon $(\D, [\ell^a])$ admits a canonical foliation
by 2-spheres $\SD$ which can be used to lift vector fields on $\PD$ to
``horizontal'' vector fields on $\D$.  In particular, then, there is a
canonical injection of the projective symmetry algebra $\psa$ into the
total symmetry algebra $\sa$.  As a result, the semi-direct sum
structure reduces simply to a direct sum structure.  In this case, one
can write \textit{any} symmetry vector $W^a$ in the form
\begin{equation}\label{decW}
	W^a = B_W\ell^a + h_W^a,
\end{equation}
where $B_W$ is a constant and $h_W^a$ is a \textit{horizontal} vector field
(i.e. tangent to the preferred 2-spheres $\SD$), satisfying $\Lie_\ell h_W^a =
0$.  The condition $\Lie_W \, \hc_a = 0$ further implies $\Lie_{h_W}\, \hc_a
\eqhat 0$, and hence $\Lie_{h_W}\, (\ImPt \Psi_2) \eqhat 0$.

\section{Angular Momentum, Mass and the First Law}
\label{s4}

In this section we will introduce definitions of mass and angular
momentum for type II weakly isolated horizons and derive the first law
of black hole mechanics in this context.  As in the non-rotating case
treated in \cite{afk}, we will use a Hamiltonian framework.

This section is divided into five parts.  In the first, we make
appropriate restrictions on the phase space to enable the introduction
of a useful notion of angular momentum.  The second sub-section
considers the issue of defining Hamiltonians generating canonical
transformations corresponding to \textit{space-time} diffeomorphisms
and provides criteria for their existence.  In the third sub-section,
we specialize this discussion to the case where the diffeomorphism
reduces to a rotational symmetry at the horizon and show the
corresponding Hamiltonian does indeed exist.  This Hamiltonian is
interpreted as the horizon angular momentum.  The fourth sub-section
considers space-time diffeomorphisms which can be interpreted as
time-translations and analyzes the issue of whether their induced
action on the phase space is Hamiltonian (i.e., preserves the
symplectic structure).  In contrast to the rotational case, the answer
is not always in the affirmative: As in \cite{afk}, the necessary and
sufficient condition for the evolution to be Hamiltonian is precisely
that the first law holds.  Thus, for every space-time vector field
$t^a$ which generates a Hamiltonian evolution, there is an associated
energy $E_\D^{(t)}$ and angular momentum $J_\D$ satisfying the first
law.  The last sub-section shows that, in the Einstein--Maxwell case,
there is a natural way to select a preferred class of evolution vector
fields $t_0^a$ for which $E_\D^{(t_0)}$ can be interpreted as the
\textit{horizon mass} $M_\D$.  The corresponding first law is then the
canonical generalization of the standard first law of black hole
mechanics to the context of rotating, weakly isolated horizons.

\subsection{The Phase Space of Rigidly Rotating Horizons}
\label{s4.1}

Physical observables such as energy and angular momentum are naturally
associated with symmetries: energy is associated with
time-translations and angular momentum with rigid rotations.  For
example, in the familiar construction of the ADM energy, one first
introduces a vector field $t^a$ in space-time which asymptotically
approaches a time-translation symmetry of the fixed flat metric at
infinity and constructs the Hamiltonian $H_t$ generating the
corresponding time-evolution in the phase space.  Since Einstein's
theory is generally covariant, ``on shell'' (i.e. when the constraints
are satisfied), the Hamiltonian reduces to 2-surface integrals on the
boundaries of the Cauchy surface under consideration.  Under the
standard assumption that there are no internal boundaries, the
on-shell value $E^{(t)}_\infty$ of the Hamiltonian is thus given just
by a 2-surface integral at infinity, which is interpreted as the total
ADM energy of space-time with respect to the asymptotic
time-translation of $t^a$.  Likewise, to define the total angular
momentum of space-time, one considers a space-time vector field
$\varphi^a$ which approaches a rigid rotation of the asymptotic metric
at infinity.  The angular momentum $J_\infty^{(\varphi)}$ is then the
surface integral at infinity giving the on-shell value of the
Hamiltonian generating rotations along $\varphi^a$.  To define energy
and angular momentum of an isolated horizon, it is therefore natural
to examine Hamiltonians which generate appropriate symmetries at the
horizon.

Let us begin with angular momentum.  In the above procedure, while
there is considerable freedom in the choice of $\varphi^a$, these
vector fields must approach a \textit{fixed} rotational Killing field
$\phi^a$ of the universal flat metric at infinity.  This condition
plays a key role in the standard proof of the existence of a
Hamiltonian generating the corresponding motions in phase space.  More
importantly, the requirement has a direct physical origin.  Angular
momentum is \textit{not} a scalar quantity in physical theories and
has several independent components.  By fixing one particular axial
symmetry across all asymptotically flat space-times, one effectively
guarantees the \textit{same component} of angular momentum is
calculated for all space-times.  If $\varphi^a$ were to approach
different rotational Killing fields of the flat background in
different space-times, even if one could construct the corresponding
conserved quantity, it would be difficult to interpret it physically.

To define angular momentum of weakly isolated horizons, it would be
natural to start with vector fields $\varphi^a$ which approach a
rotational symmetry on the horizon.  Recall however that, in contrast
to infinity, weakly isolated horizons do \textit{not} have a universal
metric and, furthermore, the metrics in class III isolated horizons
need not admit \textit{any} rotational symmetry at all.  Therefore, it
is natural first to restrict ourselves to class II, i.e.,
axi-symmetric, weakly isolated horizons.  Note however that, even in
this case, the metric on the horizon is not universal, whence a priori
we do not have a fixed rotational vector field $\phi^a$ on $\D$ that
the space-time vector fields $\varphi^a$ can be required to approach.
Therefore, for a meaningful comparison of horizon angular momenta of
different space-times, it is convenient to introduce a \textit{fixed}
rotational vector field $\phi^a$ on the horizon and admit only those
space-times in the phase space which have this $\phi^a$ as the horizon
symmetry.  We will do so.

Thus, let us now fix a vector field $\phi^a$ on the inner boundary
$\D$ of $\M$ such that: (i) it has vanishing Lie bracket with
$[\ell^a]$, (ii) vanishes on exactly two generators of $\D$, and (iii)
has closed, circular orbits of affine length $2\pi$.  For
calculational convenience, we will also insist that it be tangent to
the past boundary $\SD^-$ of the horizon.  Like the equivalence class
$[\ell^a]$, this $\phi^a$ will now be regarded as an extra structure
fixed on $\D$ once and for all.  The phase space will now consist of
the sub-manifold $\G_{\phi}$ of the covariant phase space $\G$ (of
section \ref{s2.3}), consisting of those asymptotically flat solutions
$(e^a_I, A_a{}_I{}^J, \emA_a)$ to the field equations for which $(\D,
[\ell^a], \phi^a)$ is a type II horizon with $\phi^a$ as its
rotational symmetry.  (Thus, $\Lie_{\phi} \ell^a \in [\ell^a]$,
$\Lie_{\phi} q_{ab} \eqhat 0, \Lie_{\phi} \hc_a \eqhat 0$ and
$\Lie_{\phi} F_{ab} \eqhat 0$.)  We will refer to $\G_{\phi}$ as the
phase space of rigidly rotating horizons.  In the next two
sub-sections we will show that on $\G_{\phi}$ we can use the standard
strategy of defining conserved quantities (outlined in the beginning
of this sub-section) and arrive at a definition of the horizon
angular-momentum $\JD$.

\subsection{Existence of Hamiltonian Generating Space-time 
Diffeomorphisms}
\label{s4.2}

Fix a vector field $W^a$ in each space-time in $\G_{\phi}$ such
that the diffeomorphisms it generates preserve the boundary conditions
both at spatial infinity and at the horizon.  As discussed below, the
Lie derivatives of $(e^a_I, A_a{}_I{}^J, \emA_a)$ by $W^a$ define a
vector field $\delta_W$ on $\G_{\phi}$.  The key question of this
sub-section is: Is $\delta_W$ Hamiltonian?  Or, alternatively, does
the Lie derivative of the symplectic structure along $\delta_W$
vanish?  We will find a necessary and sufficient condition for the
answers to these questions to be in the affirmative.  This result will
then be used to define the horizon angular momentum and energy in the
next two sub-sections.  In this sub-section, we will allow $W^a$ to
vary from one space-time to another.  Since we fixed the rotational
symmetry field $\phi^a$ on $\D$, this generality is not necessary for
the definition of angular momentum.  However, as we will see in
Section \ref{s4.4}, it is essential to the definition of energy
because, in contrast to spatial infinity, the near horizon geometry
can vary from one space-time to another.  This complication makes the
question of existence for certain Hamiltonians rather more subtle in
the presence of a weakly isolated horizon.

For calculational simplicity, it is useful to introduce a universal
foliation of the horizon by 2-spheres, although our final results do
not depend on it.  We saw in section \ref{s2.2} that a non-extremal,
weakly isolated horizon admits a natural foliation of this type.  But
we would like to incorporate the extremal case as well.  Let us
therefore define the leaves of the foliation as the rigid translations
of the past boundary $\SD^-$ of $\D$ along any element of the
equivalence class $[\ell^a]$.  (In the non-extremal case, these are
precisely the level surfaces of the function $\psi$ appearing in the
symplectic structure and this foliation coincides with the preferred
foliation if and only if the past boundary $\SD^-$ is a leaf of the
preferred foliation.)  The practical advantage of introducing such a
foliation is that it allows us to decompose a vector field $W^a$ on
the horizon in to vertical and horizontal components, exactly as in
\eqref{decW}.

Consider, then, a smooth assignment of a vector field $W^a$ on $\M$ to
each space-time in the phase space $\G_{\phi}$ such that, at infinity,
$W^a$ is an asymptotic symmetry, and on the horizon $W^a$ is
tangential to $\D$, and
$$W^a = B_W\ell^a + h^a_W\, ,$$ 
with $\Lie_\ell h_W^a = 0$ and $B_W$ a constant\relax
\footnote{As we saw in Section \ref{s3}, if $\kappa_{(\ell)}= 0$, 
$B_W$ need not be constant on $\D$ for $W^a$ to define a symmetry of
that horizon.  However, our purpose here is to consider a
\textit{smooth} assignment of symmetry vectors to many different
horizons.  By continuity in phase space, therefore, $B_W$ should be
constant in this case as well.}
on $\D$.  The motion in phase space associated to the diffeomorphism
along $W^a$ is given simply by the Lie derivative:
%
$$\delta_W \Sigma_{ab}{}_I{}^J = \Lie_W \Sigma_{ab}{}_I{}^J, \quad
\delta_W A_a{}_I{}^J = \Lie_W A_a{}_I{}^J \quad\mbox{and}\quad
\delta_W \emA_a = \Lie_W \emA_a. $$
%
When the background fields $(\Sigma_{ab}{}_I{}^J, A_a{}_I{}^J,
\emA_a)$ satisfy the field equations of Einstein-Maxwell theory, one
can easily verify that $\delta_W$ satisfies the linearized equations
of motion.  It therefore represents a tangent vector field on
covariant phase space.  This vector field generates a canonical
transformation if it preserves the symplectic structure, i.e., if
$\Lie_{\delta_W} \Omega = 0$.  Equivalently, $\delta_W$ is a canonical
transformation if and only if there exists a Hamiltonian function
$H_W$ on phase space such that
\begin{equation}\label{hamEq}
	\delta H_W = \Omega(\delta, \delta_W)
\end{equation}
for all tangent vectors $\delta$ to phase space.  (As with any
generally covariant theory, one expects the Hamiltonian, if it exists,
will consist only of surface terms.  We will see that this expectation
is correct.)

Using the symplectic structure \eqref{symp} and dropping terms which
vanish when the equations of motion hold and $\delta$ satisfies the
linearized equations of motion, the right side of \eqref{hamEq}
becomes
\begin{eqnarray}\nonumber
\Omega(\delta, \delta_W) &=& \frac{1}{16\pi G}\, \int_M 
\ed \Tr{\delta A \wedge (W \vins \Sigma) + (W \vins A) \delta\Sigma}
\\&&\nonumber{}
- \frac{1}{8\pi G}\, \oint_{\SD} 
(\delta\psi)\, (\delta_W \te) - (\delta_W \psi)\,  (\delta \te)
\\&&\nonumber{}
- \frac{1}{4\pi}\, \int_M 
\ed [\delta\emA \wedge (W \vins \dual\emF) + 
(W \vins \emA) \delta(\dual\emF)]
\\&&\label{ssW}{}
- \frac{1}{4\pi}\, \oint_{\SD} 
(\delta\chi)\, (\delta_W(\dual\emF)) - (\delta_W \chi)\, 
(\delta(\dual\emF)).
\end{eqnarray}
The remaining bulk terms reduce to surface integrals on $\SD$ and
$\Si$.  The integrals on $\Si$ are the usual ones and are not the main
focus of this subsection.  On the other hand, the surface terms at
$\SD$ arising from the bulk integrals, together with those already
present in \eqref{ssW}, will provide the critical test of whether
$\delta_W$ defines a symmetry of the symplectic structure; we will
focus on these.

The first issue one must address is the definition of $\delta_W
\,\psi$ and $\delta_W \,\chi$.  Since $\psi$ and $\chi$ are uniquely
determined by the triplet $(e^a_I, A_a{}_I{}^J, \emA_a)$ at the
horizon (see \eqref{grgDef} and \eqref{emgDef}), $\delta_W \,\psi$ and
$\delta_W\,\chi$ are completely unambiguous.  However, their explicit
expressions involve a subtlety.  For definiteness, let us consider
$\delta_W \psi$.  One may first be tempted to set $\delta_W \psi =
\Lie_W \psi$.  However, recall the definition \eqref{grgDef} of 
$\psi$ required $\psi$ to vanish on $\SD^-$.  The naive
definition of $\delta_W \psi$ generally does not preserve this
condition.  Hence the naive expression is incorrect and must be
modified to ensure that $\delta_W \psi$ vanishes on $\SD^-$.  To
address this problem, let us proceed systematically and return to
definition \eqref{grgDef} of $\psi$.  Using the second part to the
definition \eqref{grgDef} of $\psi$ and the properties $\Lie_W \ell^a
\eqhat 0$ and $\Lie_W \hc_a \eqhat 0$ of $W$, it follows that
$$\Lie_\ell (\delta_W \psi) \eqhat 0. $$
Since $\delta_W \psi$ must vanish on $\SD^-$, we conclude $\delta_W
\psi \eqhat 0$ on all of $\D$.  A similar argument applies to $\chi$,
whence we conclude
$$	\delta_W \psi = 0 = \delta_W \chi. $$
Note that the subtlety arose because $\psi$ and $\chi$ are
\textit{potentials} for physical fields.  Since all other terms in
\eqref{ssW} involve the fields themselves, there is no further
subtlety in defining the action of $\delta_W$; the action is given
simply by the Lie derivative.

The remainder of the calculation is straightforward.  The surface
terms in \eqref{ssW} arising from the bulk integrals can be expressed
in terms of $\hc_a$ and $\te_{ab}$.  The final result can be
conveniently expressed as
%
\begin{eqnarray}\nonumber
\Omega(\delta, \delta_W) &=& \frac{-1}{8\pi G} \oint_{\SD} 
\delta[(h_W \vins \hc) \te] - (\delta h_W \vins \hc) \te + 
\kappa_{(W)} \delta\te 
\\&&\nonumber{} 
- \frac{1}{4\pi} \oint_{\SD} \delta[(h_W \vins \emA) \dual\emF] 
- (\delta h_W \vins \emA) \dual\emF - \Phi_{(W)} \delta(\dual\emF)
\\&&\nonumber{} 
+\frac{1}{16\pi G} \oint_{\Si} \Tr{\delta A \wedge 
(W \vins \Sigma) + (W \vins A) \delta\Sigma} 
\\&&\label{dGen1}{} 
- \frac{1}{4\pi} \oint_{\Si} \delta\emA \wedge (W \vins \dual\emF) +
(W \vins \emA) \delta(\dual\emF),
\end{eqnarray}
where $\Phi_{(W)}$ denotes the electric potential of the horizon
relative to the vertical component of $W^a$.  The right side of this
result consists of integrals both at the horizon and at infinity.  If
the Hamiltonian $H_W$ is to exist, then the right side of
\eqref{dGen1} must be equal to the exact variation of some expression.
As is well-known, the surface integrals at infinity can themselves be
written as exact variations whenever $W^a$ becomes a symmetry of the
asymptotic metric at infinity.  However, there is no a priori reason
why the surface integral at the horizon is an exact variation.  Thus,
somewhat surprisingly, although the evolution generated by $W^a$ does
yield a flow on the phase space $\G_{\phi}$, \textit{this flow need
not be Hamiltonian}.  The necessary and sufficient condition for it to
be Hamiltonian is precisely that the surface integral on $\SD$ in
(\ref{dGen1}) equals $\delta H^{(W)}_\D$ for some function
$H^{(W)}_\D$ on $\G_{\phi}$ (depending only on values of horizon
fields).

\subsection{Angular Momentum}
\label{s4.3}

To define angular momentum, let us assign a vector field $\varphi^a$
on $\M$ to each space-time in $\G_{\phi}$ such that its restriction to
$\D$ is given by our fixed rotational vector field $\phi^a$.  Since we
are primarily interested only in the horizon angular momentum, to
avoid the unnecessary analysis of the terms at infinity, let us
further require that $\varphi^a$ vanishes outside some compact
neighborhood of the horizon.  (We will relax this requirement after
obtaining the expression of the horizon angular momentum.)  Then,
setting $W^a = \varphi^a$, \eqref{dGen1} simplifies: the term at
infinity vanishes and $\delta h_W^a$ can be set to zero.  Thus, we now
have
$$\Omega(\delta, \delta_{\varphi}) = 
\frac{-1}{8\pi G} \oint_{\SD} \delta[(\phi \vins \hc) \te] -
\frac{1}{4\pi} \oint_{\SD} \delta[(\phi \vins \emA) \dual\emF]. $$
The right side of this expression is clearly the variation of an
integral over $\SD$.  Therefore, we conclude the Hamiltonian
generating motions along $\varphi^a$ does indeed exist.  Moreover,
that Hamiltonian will consist of a single surface term at $\D$ whose
value may be taken to define the angular momentum $\JD$ of the
horizon:
\begin{equation}\label{jDef}
\JD := H^\phi_\D = -\frac{1}{8\pi G} \oint_{\SD} (\phi \vins \hc) \te - 
	\frac{1}{4\pi} \oint_{\SD} (\phi \vins \emA) \dual\emF,
\end{equation}
where the integral can be evaluated on \textit{any} cross-section of
$\D$.  This definition is manifestly (quasi-)local to the horizon.
Since $\hc_a$ and $\te_{ab}$ are invariant under the permissible
tetrad rotations, the gravitational term is invariant under internal
Lorentz rotations at the horizon.  Using the fact that $\Lie_\phi
(\dual\emF)_{\pback{ab}} \eqhat 0$ and the Maxwell equations, it is
easy to check that the Maxwell term is also gauge invariant.  Finally,
note that, in contrast to the standard angular momentum expressions at
infinity \cite{ah}, the horizon angular momentum $\JD$ includes
contributions from both the gravitational and electro-magnetic fields.
In this respect the right side of \eqref{jDef} is completely analogous
to the expression of the horizon energy derived in \cite{afk} in the
non-rotating case.

It is natural to ask if the horizon angular momentum can be expressed
directly in terms of space-time curvature at $\D$.  The intuition
derived from the Newman-Penrose framework suggests that the
gravitational contribution to the angular momentum is encoded in the
component $\ImPt{\Psi_2}$ of the Weyl curvature.  We will now show
that his rule of thumb is explicitly realized in the present
construction.  Since $\phi^a$ is a Killing vector of the intrinsic
horizon geometry, it is also a symmetry of the area element
$\te_{ab}$.  Thus, we find $\Lie_{\phi} \te = \ed(\phi \vins \te) =
0$, from which it follows that $\phi \vins \te = \ed f $ for some
smooth function $f$ satisfying $\Lie_{\ell} f\eqhat 0$ on the horizon.
Now, since $\phi^a$ is tangent to $\SD$, we have
%
$$\oint_{\SD} (\phi \vins \hc) \te
	= \oint_{\SD} \hc \wedge (\phi \vins \te)
		= \oint_{\SD} 2f\, \ImPt{\Psi_2} \te, $$
%
where we have performed an integration by parts in the last step.
Next, consider the electromagnetic term in \eqref{jDef}.  Since
$\Lie_{\phi}\pback{\dual\emF} \eqhat 0$ and $\ed \dual \emF \eqhat 0$,
it follows that $\phi \vins \pback{\dual\emF} = \ed g $ for some
smooth function $g$ satisfying $\Lie_\ell g \eqhat 0$ on the horizon.
Hence, using an identical argument as above we can replace the term
$(\phi \vins \emA)$ involving the electromagnetic potential by the
Newman-Penrose component $\phi_1$ of the Maxwell field (given by
$\phi_1 = -{\scriptstyle\frac{1}{4}}\,
\te^{ab} [(\dual\emF)_{ab} + i\emF_{ab}] $).  Thus, we can re-express the
horizon angular momentum \eqref{jDef} as
\begin{equation}\label{jDef2}
\JD =  -\frac{1}{4\pi G} \oint_{\SD} f\, \ImPt{\Psi_2} \te
	+ \frac{1}{2\pi} \oint_{\SD} g\, \ImPt{{\bf \phi}_1} \te.
\end{equation}
Since the integrands in \eqref{jDef2} are Lie-dragged by $\ell$, the
full expression can be projected to the 2-sphere $\hat{S}$ of
generators of $\D$.

Having the precise definition at hand, we can now ask for its relation
with other notions of angular momentum available in the literature.

First suppose that there is no Maxwell (or any other field) in a
neighborhood of $\D$ and $\varphi^a$ is in fact a Killing field of a
specific geometry in our phase-space $\G_{\phi}$ in this neighborhood.
Then, one could also define the angular momentum via Komar integral.
To analyze the relation between the two definitions, first consider
the 1-form $n_a$ on $\D$ which is normal to the foliation by $\SD$ and
is normalized with $\ell^a n_a \eqhat -1$ everywhere.  It is easy to
show $\ell^a \nabla_a n_{\pback{b}} \eqhat
\hc_b$, where $\nabla_a$ denotes the space-time connection.  Using
this relation in the gravitational contribution to the angular
momentum definition \eqref{jDef}, one finds
$$ J_\D = -\frac{1}{8\pi G}\oint_{\SD} (\phi \vins \nabla_\ell\, n) 
\te = \frac{1}{8\pi G} \oint_{\SD} (\nabla_\ell\, \phi \vins n) \te =
\frac{1}{16\pi G} \oint_{\SD} (\ell \vins \ed\varphi) \cdot n \te,$$
where we have used the Killing property of $\varphi^a$ in the last
equality.  By re-writing the last integral in terms of the the
space-time dual of the 2-form $\ed\varphi$, we obtain
$$ J_\D = \frac{1}{8\pi G} \oint_{\SD} \dual \ed \varphi\, . $$
The right side is precisely the Komar angular momentum.  Thus, in any
space-time in which $\phi^a$ can be extended to a space-time Killing
field $\varphi^a$ in a neighborhood of $\D$, the gravitational
contribution to $J_\D$ agrees with the usual Komar expression defined
by $\varphi^a$.  As at infinity, this is an exact agreement, not just
a proportionality. Finally note that, even in presence of Maxwell
fields on the horizon, the above discussion establishes the equality
of the Komar integral with the \textit{gravitational term} in
\eqref{jDef}.

A second definition of angular momentum is the one associated with
infinity \cite{ah}.  However, the integral at infinity represents the
\textit{total} angular momentum, including that in the radiation
fields outside the horizon.  Therefore, in general, one does not
expect the two to agree.  Indeed, a priori, it is not clear to
\textit{which} component of the angular momentum at infinity we should
compare $J_\D$.  However, this problem disappears if the space-time
under consideration admits a global, rotational Killing field
$\varphi^a$ (with $\Lie_{\varphi} \emF_{ab} =0$ on $\M$), whose
restriction to $\D$ is given by $\phi^a$.  Let us therefore consider
this case.

Now, since $\varphi^a$ is everywhere a Killing field of the space-time
under consideration, the vector field $\delta_{\varphi}$ on the phase
space can only define an infinitesimal gauge transformation.  However,
such a gauge transformation defines a degenerate direction of the
symplectic structure.  Thus, we have
%
$$	\delta H_\phi = \Omega(\delta, \delta_\phi) = 0 $$
%
for \textit{all} tangent vectors $\delta$ at any axially symmetric
point $(e^a_I, A_a{}_I{}^J, \emA_a)$ of the phase space.  Therefore,
on any connected component of the phase-space $\G_{\phi}$ consisting
of space-times which admit $\varphi^a$ as a Killing field, $H_{\phi}$
must be a constant.  Note, however, that with $G=c=1$, $H_{\phi}$ has
dimensions of angular momentum ((length)${}^2$) while the theory has
no dimensionful constant with that dimension (recall: the cosmological
constant is zero).  Therefore, Hamiltonian itself must vanish.
Finally, from (\ref{dGen1}) it follows that, on shell, the Hamiltonian
consists only of two surface terms:
%
$$	H_\phi = J_\D - J_\infty. $$
%
We therefore conclude: if the space-time admits a global rotational
Killing field $\varphi^a$ which reduces to $\phi^a$ on $\D$, then
\begin{equation} \label{equal}
J_\D = J_\infty
\end{equation}
where $J_\D$ is the \textit{total} horizon angular momentum (defined
by $\phi^a$).

At first this result \eqref{equal} is rather surprising, both
mathematically and physically, because one would have expected
$J_\infty$ and $J_\D$ to differ from each other by the angular
momentum in Maxwell field outside the horizon.  However, a closer
examination shows that the result is to be expected but for rather
subtle reasons.  Let us first consider the mathematical aspect.  In
the axi-symmetric case, it is well-known that $J_\infty$ is given by
the Komar integral $J_\infty^{\rm K}$ defined by $\varphi^a$
\textit{at infinity} \cite{am}.  We saw above that the gravitational
contribution to $J_\D$ equals the Komar integral $J_\D^{\rm K}$
evaluated \textit{at the horizon}.  It turns out that, using the
Cartan identity, the $\varphi^a$-angular momentum in the bulk
electromagnetic field ---i.e., the integral $\int T_{ab}\varphi^a
dS^b$ over any partial Cauchy surface extending from the horizon to
infinity--- can in fact be expressed as surface integrals.  The term
at infinity vanishes because of the fall-off conditions while the term
at the horizon is precisely the electromagnetic contribution to
$J_\D$.  Thus, the bulk electromagnetic contribution to
$\varphi$-component of angular momentum is already contained in $J_\D$
through the Maxwell horizon term.  Physically, one can understand the
situation as follows.  In general, if $\D$ extends in the future all
the way to $i^+$, the horizon angular momentum $J_\D$ is to be thought
of as the angular momentum ``left over at $i^+$ after allowing for
radiation''.  (This is completely analogous to the situation with the
horizon mass analyzed in detail in
\cite{abf}.)  Now, if $\varphi$ is a Killing vector, there is no radiation of
the $\varphi$-component of angular momentum, whence the
$\varphi$-component evaluated at $i^+$ is the same as that evaluated
at $i^o$, whence $J_\D = J_\infty$.

We will conclude with three remarks.

1. Let us restrict ourselves to class I (i.e. spherically symmetric)
weakly isolated horizons.  In this case, $q_{ab}, \hc_a$ and $\emA_a$
are Lie-dragged by all three rotational vector fields.  Hence, the
pull-backs to the spherical sections of $\hc_a$ and $\emA_a$ must
vanish and the integrand of \eqref{jDef} therefore vanishes.  Thus, as
one would physically expect, $J_\D$ vanishes on all class I horizons.

2.  The vector field $\delta_{\varphi}$ determines the Hamiltonian
$H_{\varphi}$ only up to an additive constant.  How was this freedom
fixed in \eqref{jDef}?  In the case where the cosmological constant
vanishes, this constant can be fixed to zero because there is no
parameter available in the theory with the correct dimension.  Even
when the cosmological constant is not zero, the freedom can still be
eliminated, e.g., simply by the physical requirement that horizon
angular momentum should vanish in class I. As we just showed, $J_\D$
satisfies this condition.  Thus, the additive constant is fixed in a
natural fashion.

3.  An important question for this framework is whether the event
horizon of a Kerr--Newman black hole is a rigidly rotating isolated
horizon and, if so, whether the above definition of angular momentum
reproduces the standard result.  The answer, in both cases, is in the
affirmative.  Being Killing horizons, these event horizons are in
particular weakly isolated horizons and the axi-symmetry of the
ambient space-time defines the field $\phi^a$ on $\D$.  Thus, these
are rigidly rotating isolated horizons.  Finally, the discussion above
shows that in this case $J_\D$ equals $Ma$, the standard angular
momentum $J_\infty$ defined at infinity.  Thus (as with the horizon
mass in the non-rotating case \cite{afk}), our horizon angular
momentum expression \eqref{jDef} contains not only the ``bare''
angular momentum one may naively associate with the horizon but also
the contributions from the electromagnetic hair outside.

\subsection{The First Law}
\label{s4.4}

To state the first law of black hole mechanics for isolated horizons,
we must first define horizon energy.  As with angular momentum in the
previous sub-section, we will base our definition of energy on the
Hamiltonian generating translations along an appropriate symmetry of
the horizon.  However, which particular symmetry $W^a = t^a$ would
correspond to the desired time-translations is not immediately clear.
Therefore, we will begin with an \textit{arbitrary} symmetry field
$t^a$.  At the horizon, then, it only has to satisfy
\begin{equation} \label{sym}
t^a + \Omega_{(t)} \phi^a \in [\ell^a]\quad {\rm or} \quad
t^a = B_{(\ell, t)} \ell^a - \Omega_{(t)} \phi^a
\end{equation}
for some \textit{constants} $\Omega_{(t)},\, B_{(\ell, t)}$ on $\D$.
(The latter depends not only on the specific choice of $t^a$ but also
of $\ell^a \in [\ell^a]$.)  The constant $\Omega_{(t)}$ will be
referred to as the \textit{angular velocity of the horizon relative
to} $t^a$.  Note that both the specific element $B_{(\ell,t)}\ell^a$
of $[\ell^a]$ and $\Omega_{(t)}$, determined by $t^a$, can depend on
the dynamical fields $(e^a_I, A_a{}_I{}^J, \emA_a)$; in the numerical
relativity terminology, we are considering `live'' evolution vector
fields $t^a$.  This generality is essential in particular because the
physically appropriate angular velocity $\Omega_{(t)}$ of $\D$ will
vary from one space-time to another.  To summarize, to each point in
the phase space $\G_\phi$, we assign a vector field $t^a$ on $\M$
satisfying
\eqref{sym} on $\D$, allowing the vector field to vary from one point of the
phase space to another.

To analyze the question of whether $\delta_t$ is a Hamiltonian vector
field on $\G_\phi$, we must determine whether the right side of
Hamilton's equations \eqref{hamEq} is the exact variation of some
Hamiltonian.  We have already calculated the quantity in question for
an arbitrary vector field $W^a$ in \eqref{dGen1}. Since the surface
terms at infinity are not central to this discussion, as in the
angular momentum calculation above, for simplicity let us first assume
that $t^a$ vanishes outside of some compact neighborhood of $\D$.  The
only potential obstruction to $\delta_t$ being Hamiltonian lies in the
horizon surface term of \eqref{dGen1}.  Being linear in $\delta_t$,
this surface term defines a 1-form $X_\D^t$ on phase space. From the
right side of \eqref{dGen1}, $X_\D^t$ can be expressed as
\begin{eqnarray}\nonumber
X_\D^t(\delta) &=& \frac{1}{8\pi G} \oint_{\SD} 
\delta[(-\Omega_{(t)} \phi \vins \hc) \te] - 
[\delta(-\Omega_{(t)} \phi) \vins \hc] \te + \kappa_{(t)} \delta\te 
\\&&\qquad\label{Xdef}{}
+ \frac{1}{4\pi} \oint_{\SD} \delta[(-\Omega_{(t)} \phi 
\vins \emA) \dual\emF] - 
[\delta(-\Omega_{(t)} \phi) \vins \emA] \dual\emF + 
\Phi_{(t)} \delta(\dual\emF)
\end{eqnarray}
where we have decomposed $t$ as in \eqref{decW} with $h_t =
-\Omega_{(t)}\, \phi^a$. Let us begin with the first integral.  Since
$\phi$ is fixed once and for all on $\D$, the variation $\delta$ in
the second term only affects $\Omega_{(t)}$.  Therefore, the second
term will simply cancel the part of the first term where the variation
hits $\Omega_{(t)}$ and we can therefore write the first two terms
together as
%
$$\frac{1}{8\pi G} \oint_{\SD} -\Omega_{(t)} 
\delta[(\phi \vins \hc) \te]. $$
%
However, the quantity $\Omega_{(t)}$ is constant over $\D$ and may 
therefore be pulled outside the integral.  The remaining integrand is 
exactly the $\delta$-variation of the gravitational contribution to the 
angular momentum \eqref{jDef}.  Analogous procedure can be applied to the 
second (i.e., electro-magnetic) term of \eqref{Xdef}, likewise yielding the 
electro-magnetic contribution to $\JD$.  Furthermore, since the surface 
gravity and electro-static potential are both constant over $\D$, we can 
pull them outside the respective integrals and \eqref{Xdef} reduces to
\begin{equation}\label{Xval}
	X_\D^t(\delta) = \frac{\kappa_{(t)}}{8\pi G}\, \delta\AD + 
	\Omega_{(t)} \delta \JD + \Phi_{(t)} \delta \emQD.
\end{equation}
Note that the right side of \eqref{Xval} is strongly reminiscent of
the first law of black hole mechanics. However, at this stage of our
analysis, it only provides an explicit expression of the 1-form
$X_\D^t$ on $\G_\phi$.  The condition that $\delta_t$ be Hamiltonian is 
simply that the 1-form $X_\D^t$ is closed:
\def\wwedge{\mathbin{\hbox{\rlap{$\mathalpha\wedge$}$\,\mathalpha\wedge$}}}
\begin{equation}\label{dX}
0 = \dd X_\D^t 
= \frac{1}{8\pi G}\, \dd\kappa_{(t)} \wwedge \dd\AD 
+ \dd\Omega_{(t)} \wwedge \dd\JD
+ \dd\Phi_{(t)} \wwedge \dd\emQD,
\end{equation}
where $\dd$ and $\wwedge$ denote the exterior derivative and exterior
product on the (infinite-dimensional) phase space $\G_\phi$.  This
simple relation leads to some startling consequences which we now
discuss.

A priori, the horizon value of $t^a$ can vary from one space-time to
another in any smooth fashion; for each such choice, we obtain a flow
on the phase space.  Eq \eqref{dX} implies that most of these flows
fail to preserve the symplectic structure.  To begin with, for the
flow to be Hamiltonian, the surface gravity $\kappa_{(t)}$, the
angular velocity $\Omega_{(t)}$ and the electric potential
$\Phi_{(t)}$ can be functions \textit{only of the area $a_\D$, angular
momentum $J_\D$ and the charge $\emQD$}. 
\footnote{So far, our construction is very general. For any given 
physical metric (satisfying our boundary conditions) we can choose an
$[\ell]$ so that $(\D, [\ell])$ is an isolated horizon with
$\kappa_{(\ell)} =0$ \textit{and} another equivalence class $[\ell']$
with $\kappa_{(\ell')} \not=0$. Hence, for any values of the triplet
$(\AD, \JD, \emQD)$, and an evolution field $t^a \eqhat B\ell^a - 
\Omega\phi^a$ (with $B \not= 0$), there is a phase space point with
$\kappa_{(t)}\equiv B \kappa_{(\ell)} = 0$ and another with
$\kappa_{(t)} \not= 0$. Therefore, irrespective of the choice of
$t^a$, the evolution will fail to be Hamiltonian unless we remove the
spurious redundancy in the phase space. This can be naturally
accomplished as follows.  Denote by $\kappa_{\rm KN}(\AD,\JD,\emQD)$
the function of the three intrinsic horizon parameters which yields
surface gravity in the Kerr--Newman family.  We will excise those
points from our phase space for which one of $(\kappa_{(\ell)},
\kappa_{\rm KN})$ vanishes but the other does not. This excision could
have been carried already in Section \ref{s2.3} when we introduced the
phase space. We did not do so because the reason behind the excision
becomes clear only after \eqref{dX}. Note that the phase space
continues to contain all Kerr--Newman solutions, including the extremal
ones, after the excision.}
Other factors such as the local geometry (i.e. distortion) of the
horizon can not affect these ``extrinsic parameters''.  Moreover,
these parameters must satisfy certain non-trivial relations:
\begin{eqnarray}\nonumber
\frac{\partial \kappa_{(t)}(\AD, \JD, \emQD)}{\partial \JD} &=& 
8\pi G\, \frac{\partial \Omega_{(t)}(\AD, \JD, \emQD)}{\partial \AD} 
\\ \nonumber{}
\frac{\partial \kappa_{(t)}(\AD, \JD, \emQD)}{\partial \emQD} &=&
8\pi G\, \frac{\partial \Phi_{(t)}(\AD, \JD, \emQD)}{\partial \AD} 
\\ \nonumber
\frac{\partial \Omega_{(t)}(\AD, \JD, \emQD)}{\partial \emQD} &=& 
\frac{\partial \Phi_{(t)}(\AD, \JD, \emQD)}{\partial \JD} \\
 \label{mixP}
\end{eqnarray}
Now, on any given point $(e^a_I, A_a{}_I{}^J, \emA_a)$ in the phase
space, $\kappa_{(t)},\, \Omega_{(t)}$ and $\Phi_{(t)}$ are completely
determined by the horizon value of $t^a$.  Conversely, at any given
phase space point, $\kappa_{(t)}$ uniquely determines the vertical
component of $t^a$ and $\Omega_{(t)}$, the horizontal component.
Therefore, relations \eqref{mixP} constrain the permissible choices of
the assignment of $t^a$ to each space-time in $\G_\phi$.  These strong
restrictions are the necessary and sufficient conditions for the flow
generated by $\delta_t$ on $\G_\phi$ to be Hamiltonian.

Thus, the restriction to the horizon of a permissible evolution field
$t^a$ is determined by this remarkably small set of parameters of the
horizon.  Note, however, that these arguments do not provide a
\textit{specific} choice of functions $\kappa_{(t)}$ and
$\Omega_{(t)}$ of the horizon parameters $\AD, \JD, {\emQD}$.
Correspondingly, for \textit{any} choice of these functions satisfying
\eqref{mixP}, the Hamiltonian generating evolution along the
corresponding $t^a$ is guaranteed to exist. (The term at infinity is
already an exact variation and is therefore not relevant to our
discussion.)  The horizon surface term in that Hamiltonian is a
natural measure of the \textit{energy} $E_\Delta^t$ of the horizon
relative to the evolution field $t^a$.  By virtue of \eqref{Xdef} and
the calculations above, the energy $E^t_\D$ is a function only of
$\AD$, $\JD$ $\emQD$ and satisfies
\begin{equation}\label{gLaw1}
\delta E_\D^t = \frac{\kappa_{(t)}}{8\pi G}\, \delta\AD 
+ \Omega_{(t)}\, \delta\JD + \Phi_{(t)}\, \delta\emQD.
\end{equation}
This is our \textit{generalized first law}, which holds for all weakly 
isolated horizons.  The analogy to the usual first law of black hole 
mechanics is clear. 

Let us summarize. A priori there is freedom to assign an evolution
vector field $t^a$ on $\M$ to every point in the phase space $\G_\phi$
in \textit{any} smooth fashion; each assignment provides an evolution
flow $\delta_t$ on $\G_\phi$.  However, most of these flows fail to
preserve the symplectic structure.  They do so if and only if the
assignment is such that \eqref{dX} (or, equivalently, \eqref{mixP})
holds.  While this is a severe restriction on the assignment of $t^a$,
as discussed below, it still leaves considerable freedom in the choice
of the assignment.  For each such ``permissible'' $t^a$, there is a
well-defined horizon energy and the first law holds.  Thus, there is a
precise sense in which the first law \eqref{gLaw1} is a necessary and
sufficient condition for the $\delta_t$-evolution to be Hamiltonian.

Since there is generally no canonical choice of a single, ``correct''
evolution field at the horizon, there is no canonical notion of the
horizon energy.  While $E^t_\D$ has a direct Hamiltonian
interpretation in the phase space $\G_\phi$, for a general permissible
$t^a$ it does not admit an obvious \textit{space-time} interpretation.
In the next sub-section, we will show that a canonical choice of $t^a$
can be made using no-hair theorems in Einstein-Maxwell theory.  The
corresponding $E^t_\D$ can be interpreted as the horizon mass.

We will conclude with two remarks.

1. There is a constructive procedure to obtain permissible vector
fields.  Choose any smooth function $\kappa_0$ of $\AD,\JD,\emQD$
satisfying the following regularity condition for each choice of
$\JD,\emQD$: the integrals
$$ \int_{a_0}^\infty \ed \AD\,\, \frac{\partial \kappa_{0}}
{\partial \JD}  \quad {\rm and} \quad \int_{a_0}^\infty \ed 
\AD\,\, \frac{\partial \kappa_{0}}{\partial \emQD}  $$
converge to well-defined functions of $\JD$ and $\emQD$, with $a_0=
4\pi\sqrt{Q^4 + 4 J^2}$ (and $\kappa_0$ vanishes if and only if 
$\kappa_{\rm KN}$ vanishes). Then, integrating the first of Eqs
\eqref{mixP} with respect to $\AD$ and requiring $\Omega
(\AD,\JD,\emQD)$ to tend to zero as $\AD$ tends to infinity for any
fixed values of $\JD, \emQD$, we obtain a unique function
$\Omega_0(\AD,\JD,\emQD)$.  Now, given any point $(e^a_I, A_a{}_I{}^J,
\emA_a)$ in $\G_\phi$, there is a unique vector field $t^a$ on $\D$
such that $\kappa_{(t)} = \kappa_0$ and $\Omega_{(t)} = \Omega_0$ 
(namely, $t^a \eqhat B_0 \ell^a + \Omega_0 \phi^a$ where $B_0$ is
given by $\kappa_0 = B_0 \kappa_{(\ell)}$).  Finally, we can integrate
the second of Eqs \eqref{mixP} with respect to $\AD$ and require that
$\Phi_{(t)}$ tends to zero as $\AD$ tends to infinity (keeping $\JD$
and $\emQD$ fixed), we obtain a function $\Phi_{(t)}$ of
$\AD,\JD,\emQD$ which satisfies the second and the third of Eqs
\eqref{mixP}.  Thus, given a sufficiently regular function $\kappa_0$
of $\AD,\JD,\emQD$, we can integrate \eqref{mixP} and, using
physically motivated conditions to determine integration constants,
obtain an admissible evolution field $t^a$ and electric potential
$\Phi_{(t)}$ on $\D$.

2. If we now drop the restriction that $t^a$ vanish outside a finite
neighborhood of $\D$ in $\M$ but require instead that it
asymptotically approach a time-translation Killing field of the fixed
flat metric at infinity, we find that the total Hamiltonian has two
terms:
$$ H_t = E^t_{\infty} - E^t_\D $$
where $E^t_{\infty}$ is the ADM energy.  Again, general arguments from
symplectic geometry imply that $H_t$ is constant on each connected
component of stationary solutions, assuming of course that $t^a$
coincides with the stationary Killing field in these space-times.  As
in the case of angular momentum, we can argue that the value of this
constant can only be zero in the Einstein--Maxwell theory.  Thus, in
each stationary space-time, the horizon energy $E^t_\D$ equals the ADM
energy at infinity.

\subsection{Horizon Mass}
\label{s4.5}

The procedure that led us to the definition of the horizon energy
$E^t_\D$ is the same as the one used at spatial infinity to define the
ADM energy.  However, the boundary conditions at infinity are such
that the asymptotic value of the evolution field $t^a$ must coincide
with one of the time translation Killing fields of the fixed flat
metric, whence the space of viable time-translations at infinity is
three-dimensional (corresponding to the unit space-like hyperboloid in
the tangent space of $i^o$).  On the other hand, at the inner isolated
horizon boundary, the physically appropriate $t^a$ necessarily varies
from one space-time to another.  In particular, if the horizon is
non-rotating, $t^a$ points along the null normal to the horizon while
if it is rotating, it has a component also along the rotational
symmetry $\phi^a$.

In the usual treatments of the first law, one restricts oneself to
perturbations of stationary backgrounds.  Consequently, one can single
out a preferred time translation from the 3-parameter family, adapted
to the rest frame of the black hole.  The corresponding ADM energy is
then also the ADM mass.  It is natural to ask if we can similarly
single out a canonical time translation also at the horizon and
introduce the notion of horizon mass.  Note that this task is
significantly more difficult than the corresponding task at spatial
infinity first because the permissible time translations at the
horizon form an \textit{infinite} dimensional family rather than three
and second because, on physical grounds mentioned above, this
canonical time translation at the horizon must vary from one
space-time to another.  Nonetheless, because of the remarkably strong
restrictions on the extrinsic parameters $\kappa_{(t)},\,
\Omega_{(t)}$ and $\Phi_{(t)}$ and thanks to the no-hair theorems,
there is a natural solution to this problem.

We have just seen in the previous sub-section that the restriction of
\textit{any} evolution vector field $t^a$ to the horizon is determined
by $\kappa_{(t)}$ and $\Omega_{(t)}$ and if the $t^a$-evolution is to
define a Hamiltonian flow on $\G_\phi$, $\kappa_{(t)}$ and
$\Omega_{(t)}$ can only be functions of $\AD, \JD$ and $\emQD$.  These
three quantities may be regarded as the independent parameters of the
horizon.  The problem of defining a canonical time-translation on the
horizon therefore reduces to that of making a canonical choice of the
two functions $\kappa_{(t)}(\AD,\JD,\emQD)$ and $\Omega_{(t)}(\AD,
\JD, \emQD)$.  Now, event horizon of a Kerr--Newman black hole is in
particular a weakly isolated horizon.  It is natural, on these
stationary space-times, to choose the evolution field to coincide with
the stationary Killing field.  This choice selects for us specific
functions of the three parameters:
\begin{equation} \label{fix}
\def\w{\sqrt{(\RD^2 + G\emQD^2)^2 + 4G^2\JD^2}}
\kappa = \frac{\RD^4 - G^2(\emQD^4 + 4\JD^2)}
{2\RD^3 \w} \quad\mbox{and}\quad \Omega = 
\frac{2G\JD}{\RD\w}\, ,
\end{equation}
where, as before, $a_\D = 4\pi \RD^2$.  Now, although the choice
\eqref{fix} was made \textit{only} for stationary cases, since there
is exactly one Kerr--Newman solution for each set of isolated horizon
parameters, this choice uniquely fixes the two functions.  Thus,
\textit{we can select a canonical time-translation on any isolated
horizon by requiring that the surface gravity and the angular velocity
it defines be given by} \eqref{fix}.  (This procedure is unambiguous
because $\RD, \JD,\emQD$ are determined by $(e^a_I, A_a{}_I{}^J,
\emA_a)$ without any reference to $t^a$.)  We will make this choice
and denote by $t^a_0$ any time-translation whose restriction to the
horizon coincides with this canonical choice.  The conditions
\eqref{mixP} then require us to partially gauge fix the Maxwell field
such that
\begin{equation}
\Phi = \frac{\emQD}{\RD}\, \frac{\RD^2 + G\emQD^2}
{\sqrt{(\RD^2 + G\emQD^2)^2 + 4G^2\JD^2}}.
\end{equation}
(Again this is the same value which the electro-static potential of
the horizon takes in a Kerr--Newman solution.) The evolution generated
by such a \textit{live} vector field $t_0^a$ is then guaranteed to be
Hamiltonian.  We can then ``integrate'' the expression \eqref{gLaw1}
of the first law to find the horizon energy $E^t_\D$.  Setting it
equal to the mass $M_\D$, we find:
\begin{equation} \label{mass}
\MD = \frac{\sqrt{(\RD^2 + G\emQD^2)^2 + 4G^2\JD^2}}
{2GR_\Delta}.
\end{equation}
Again, this gives the usual ADM mass for the Kerr--Newman solutions.
However, this definition applies to \textit{all} isolated horizons
including those which admit radiation in the exterior region.
Therefore, in general, $M_\D$ differs from the ADM mass $M_{\rm ADM}$
due to the energy in the radiation field.  In the Einstein-Maxwell
theory under consideration, if the weakly isolated horizon were to
extend all the way to $i^+$ in the distant future, $M_\D$ can be
thought of as the future limit of the Bondi mass at $i^+$ \cite{abf}.
If the space-time is globally stationary, there is no flux of
radiation across future null infinity ${\cal I}^+$ whence the future
limit of the Bondi mass coincides with the ADM mass. Finally, we wish
to emphasize that we did not simply postulate \eqref{mass}; it was
systematically derived using Hamiltonian methods.
 
With this (quasi-)local definition of the horizon mass, for any
preferred time-translation $t^a_0$, we have a \textit{canonical}
generalized first law
\begin{equation}\label{gLaw2}
\delta M_\Delta = \frac{\kappa}{8\pi G} \,\delta\AD + 
\Omega\, \delta\JD + \Phi\,  \delta\emQD\, ,
\end{equation}
applicable to all weakly isolated horizons in Einstein--Maxwell theory.

\section{Discussion}
\label{s5}

In this paper, we gave a definition of angular momentum $J_\D$ for all
rigidly rotating isolated horizons. The definition is (quasi-)local to
the horizon; it makes no reference to infinity at all. In contrast to
the usual angular momentum expression \cite{ah} at infinity, $J_\D$
has an explicit contribution from the Maxwell field on the horizon.
If the space-time in a neighborhood of the horizon is axi-symmetric
and the matter fields vanish in that region, $J_\D$ equals the usual
Komar integral formula, evaluated at the horizon. If the space-time is
globally axi-symmetric, irrespective of whether there is Maxwell field
on the horizon, $J_\D$ equals the total angular momentum at
infinity. If the weakly isolated horizon extends all the way to $i^+$,
we can regard $J_\D$ as the future limit of angular momentum along
future null infinity ${\cal I}^+$. While these properties of the
$J_\D$ are very similar to those of horizon mass $M_\D$, there is also
an interesting difference: Whereas in presence of radiation in the
exterior region $M_\D$ is always different from $M_{\rm ADM}$, if the
radiation field is axi-symmetric, $J_\D$ equals the angular momentum
at infinity (along the rotational Killing vector).

We have also generalized the standard first law of black hole
mechanics.  Although the final form \eqref{gLaw2} of this law is
identical to that of the standard one, there are some important
differences. First, our law is applicable to all space-times which
admit an isolated horizon inner boundary, including those which allow
radiation arbitrarily close to the horizon.  Second, all quantities
and variations that enter the first law are defined \textit{locally at
the horizon}. In standard treatments, the physical meaning and
appropriateness of variations is not as clear because some quantities
such as area, surface gravity and the angular velocity of the horizon
are defined at the horizon while others, like energy, angular momentum
(and sometimes even the electric scalar potential) are evaluated at
infinity.  Third, other treatments based on a Hamiltonian framework
\cite{review} often critically use the bifurcate 2-surface which does
not exist in the extremal case.  Therefore, extremal black holes are
often excluded from the first law.  The present analysis never makes
reference to bifurcate surfaces (which do not exist in physical
space-times resulting from gravitational collapse). Therefore, our
discussion of the first law holds also in the extremal case. Finally,
with obvious modifications of boundary conditions at infinity, our
analysis includes cosmological horizons where thermodynamic
considerations are also applicable \cite{gh}.

Perhaps the most important aspect of this analysis is that it sheds
new light on the `origin' of the first law: as in the non-rotating
case treated in \cite{afk}, it arose as a necessary and sufficient
condition for the existence of a Hamiltonian generating time
evolution.  A new feature of our framework is the existence of an
infinite family of first laws corresponding to the infinite family of
evolution ``permissible'' vector fields $t^a$ (i.e., vector fields
satisfying the necessary and sufficient condition \eqref{dX} for the
evolution to be Hamiltonian.)  In the Einstein--Maxwell case, using
the no-hair results, one can select a canonical ``live'' evolution
field $t_o^a$.  Correspondingly, there is a canonical notion of energy
which can be interpreted as the horizon mass $M_\D$ and hence a
canonical first law.  In more general theories which allow hairy black
holes, a canonical horizon mass can not be defined on the full phase
space.  Yet, even in this case, the isolated horizon framework is
directly useful: it enables one to relate properties of these hairy
black holes to those of the corresponding solitons
\cite{acs}.

Finally, our Lagrangian and Hamiltonian frameworks are based on real
tetrads and Lorentz (rather than self-dual) connections. It is
therefore quite straightforward to extend our analysis to any
space-time dimension.

\section*{Acknowledgments}

We would like to thank Alejandro Corichi, Olaf Dreyer, Steve
Fairhurst, Sean Hayward, Badri Krishnan, Daniel Sudarsky, Bob Wald and
Jacek Wisniewski for discussions.  This work was supported in part by
the NSF grants PHY00-90091, INT97-225514, PHY97-34871, the Polish
Committee for Scientific Research grant 2 P03B 060, the Eberly
research funds of Penn State and the Albert Einstein Institute of the
Max Planck Society. CB was supported in part by a Braddock Fellowship.

\appendix

\section{General Hamiltonians}
\label{a:diff}

In section \ref{s4}, we constructed Hamiltonians generating the
infinitesimal symmetries of rigidly rotating weakly isolated horizons.
While this is the most interesting case from a physical perspective,
for completeness, one may also wish to consider more general
space-time vector fields $W^a$ which are tangential to $\D$ and
preserve the fixed equivalence class $[\ell^a]$ thereon and ask: Can
any of these lead to Hamiltonian evolutions on the \textit{full} phase
space $\G$?  In this appendix we will analyze this issue.  While a
general $W^a$ will of course not lead to a Hamiltonian evolution,
there is an interesting sub-class which does.  It may be useful in
future investigations.

Let us begin with an assignment of a vector field $W^a$ to each
space-time in the full phase space $\G$ which preserves $(\D,
[\ell^a])$ but is not necessarily an infinitesimal symmetry in the
sense of Section \ref{s4} (i.e., does not necessarily preserve
$(q_{ab}, \hc_a, \emA_{a})$).  As in Section
\ref{s4}, we can decompose $W^a$ into vertical and horizontal parts:
$$ W^a = B_W \ell^a + h_W^a \, $$
with $\Lie_{\ell}\, h_W \eqhat 0$.  Note that the condition
$\Lie_{W}\, \ell^a \in [\ell^a]$ implies $B_W$ is constrained only by
$\Lie_\ell B_W \eqhat 0$.  The variations $\delta_W\, \psi$ and
$\delta_W\, \chi$ of the potentials become
\begin{equation}
  \delta_W \psi = (B_W - B^{-}_W) \kappa_{(\ell)},\quad {\rm and}
  \quad \delta_W \chi = (B_W - B^{-}_W) \Phi_{(\ell)}\, 
\end{equation}
where $B_W^-$ denotes the restriction of $B_W$ to the past boundary
$\SD^-$ of the horizon.

We can now calculate the action of the symplectic structure on the
vector field $\delta_W$ on $\G$.  The derivation of \eqref{ssW} from
\eqref{symp} did not depend on any way on $W^a$ defining a symmetry of
the horizon.  We can therefore begin at that point and substitute the
new definitions of $\delta_W \psi$ and $\delta_W \chi$ to find the
analog of \eqref{dGen1} for generic horizon diffeomorphisms.  If fact,
the result will be quite similar: For any tangent vector field
$\delta$ on $\G$, we will have
\begin{eqnarray}\nonumber
  \Omega(\delta, \delta_W) &=& \frac{-1}{8\pi G} \oint_{\SD} 
   \delta[(h_W \vins \hc) \te] - (\delta h_W \vins \hc) \te + 
    \kappa_{(W^-)} \delta\te 
\\&&\nonumber{}
- \frac{1}{4\pi} \oint_{\SD} \delta[(h_W \vins \emA) \dual\emF] - 
(\delta  h_W \vins \emA) \dual\emF - \Phi_{(W^-)} \delta(\dual\emF) 
\\&&\nonumber{} + \frac{1}{16\pi G} \oint_{\Si} \Tr{\delta A \wedge (W 
\vins \Sigma) + (W \vins A) \delta\Sigma} \\&&\label{dGen2}{} - 
\frac{1}{4\pi} \oint_{\Si} \delta\emA \wedge (W \vins \dual\emF) + (W \vins 
\emA) \delta(\dual\emF).
\end{eqnarray}
The only difference here is that $\kappa_{(W^-)}$ denotes the surface
gravity relative to the vertical part $B_W \ell^a$ of $W^a$,
\textit{evaluated at $\SD^-$}, and similarly for the electric
potential $\Phi_{(W^-)}$.  (On other sections $\SD$ of the horizon,
the surface gravity and electric potential may take other values.)
Since the right side contains several terms which are not exact
variations, there are a number of possible barriers to the existence
of a Hamiltonian generating motions along $W^a$.  However, as we will
now show, there are a couple of interesting cases where the required
Hamiltonian does exists.

Consider first the case when $W$ is purely horizontal at the horizon, $W^a
\eqhat h_W^a $, \textit{with, moreover, $h_W^a$ fixed} (i.e., independent of
the space-time under consideration).  Thus, $W^a$ generates a fixed
diffeomorphism on $\D$ which preserves the family of cross-sections
$S_\D$ which, however, need not be a rotation.  Since we are
interested in the horizon structure, let us also assume that $W^a$
vanishes in a neighborhood of infinity.  Then, the integrals at
infinity in \eqref{dGen2} vanish and only the first two terms in each
of the integrals at $\SD$ survive.  Moreover, the second term in each
integral vanishes because $\delta h_W^a$ is zero.  Thus, the flow
generated by these $W^a$ preserves the symplectic structure and the
corresponding Hamiltonian is given by
\begin{equation} \label{genj}
    H_{h_W} := \frac{-1}{8\pi G} \oint_{\SD} (h_W 
\vins \hc) \te - \frac{1}{4\pi} \oint_{\SD} (h_W \vins \emA) \dual\emF.
\end{equation}
Note that this integral has the same form as the integral \eqref{jDef}
used to define the angular momentum in section \ref{s4}.  However,
this formula applies to a general weakly isolated horizon, which need
not be rigidly rotating.  Moreover, the horizontal vector field
$h_W^a$ is arbitrary; it need not be a Killing vector of $q_{ab}$ and
indeed it need not even preserve the area-form $\te$.  (Consequently,
we can not re-express \eqref{genj} in terms of curvatures as in
\eqref{jDef2}.)  One might hope this formula could be used to define
angular momentum in some sense for generic weakly isolated horizons.
However, while the phase space interpretation of \eqref{genj} is
clear, if $h_W^a$ does not generate rotations, its space-time
interpretation is quite obscure.

Finally there is another family of vector fields $W^a$ for which the
flow $\delta_W$ on $\G$ is Hamiltonian, although physically less
interesting.  Let $W^a$ be purely vertical at the horizon, $W^a \eqhat
B_W\ell^a$, with $B^{-}_W \eqhat 0$.  As before, since we want to focus
on the horizon terms, let us assume that $W^a$ vanishes outside some
neighborhood of $\D$.  Then, terms at infinity in \eqref{dGen2} vanish
and because $h_W^a \eqhat 0$, the first two terms in each of the
surface integrals at $S_\D$ also vanish.  Finally, since $B^{-}_W
\eqhat 0$, these remaining terms also vanish.  Thus, the entire right
side of \ref{dGen2} vanishes.  Consequently, the associated $\delta_W$
is a \textit{degenerate direction} of the symplectic structure,
whence, from the perspective of the Hamiltonian framework, $\delta_W$
generates a \textit{gauge transformation}.  Such transformations have
no direct physical interest and may be quotiented out of the algebra
of kinematically allowable diffeomorphisms.  Indeed, since any $W^a$
of the form described above can be written uniquely as a combination
of one of these gauge transformations and another kinematically
allowable vector field with $B_W$ \textit{constant} over $\D$, we may
restrict attention to the later case without any loss of generality.

\end{document}